\documentclass[letterpaper,useAMS,usenatbib]{mn2e}
\pdfoutput=1
\usepackage{xspace}
\usepackage{amssymb}
\usepackage{graphicx} 
\usepackage{amsmath}
\usepackage[T1]{fontenc}

\newcommand{\msun}{M$_{\odot}$\xspace} 
\newcommand{\dnu}{$\Delta \nu$\xspace} 
\newcommand{\numax}{$\nu_{\mathrm{max}}$\xspace} 
\newcommand{\teff}{$T_{\mathrm{eff}}$\xspace} 
\newcommand{\aFe}{[$\alpha$/M]\xspace} 
\newcommand{\al}{$\alpha$} 
\usepackage[normalem]{ulem}

\title{Red giant masses and ages derived from carbon and nitrogen abundances}

\author[M. Martig et al.]{Marie Martig$^{1}$, Morgan Fouesneau$^{1}$, Hans-Walter Rix$^{1}$, Melissa Ness$^{1}$, 
\newauthor Szabolcs M{\'e}sz{\'a}ros$^{2}$, D. A.  Garc\'{\i}a-Hern\'{a}ndez$^{3,4}$, Marc Pinsonneault$^5$,
\newauthor Aldo Serenelli$^6$, Victor Silva Aguirre$^7$, Olga Zamora$^{3,4}$\\
$^1$Max-Planck-Institut f\"{u}r Astronomie, K\"{o}nigstuhl 17, 69117 Heidelberg, Germany\\
$^2$ELTE Gothard Astrophysical Observatory, H-9704 Szombathely, Szent Imre Herceg st. 112, Hungary\\
$^3$Instituto de Astrof\'{\i}sica de Canarias (IAC), V\'{\i}a Lactea s/n, E-38200 La Laguna, Tenerife, Spain\\ 
$^4$Departamento de Astrof\'{\i}sica, Universidad de La Laguna (ULL),E-38206 La Laguna, Tenerife, Spain\\
$^5$Ohio State University, Dept. of Astronomy, 140 W. 18th Ave., Columbus, OH 43210, USA\\
$^6$Instituto de Ciencias del Espacio (ICE/CSIC-IEEC) Campus UAB, Carrer de Can Magrans S/N Cerdanyola del Valles, 08193 Spain\\
$^7$Stellar Astrophysics Centre, Department of Physics and Astronomy, Aarhus University, Ny Munkegade 120, DK-8000 Aarhus C, Denmark\\
}
\begin{document}
\maketitle
\begin{abstract} 
We show that the masses of red giant stars can be well predicted 
from their photospheric carbon and nitrogen abundances, in conjunction with their spectroscopic stellar labels log~$g$, \teff, and [Fe/H]. This is qualitatively expected from mass-dependent post main sequence evolution. We here establish an empirical relation between these quantities by drawing on 1,475 red giants with asteroseismic mass estimates from \textit{Kepler} that also have spectroscopic labels from APOGEE DR12. We assess the accuracy of our model, and find that it predicts stellar masses with fractional r.m.s. errors of about 14\% (typically 0.2 \msun). From these masses, we derive ages with r.m.s errors of 40\%. 
This empirical model allows us for the first time to make 
age determinations (in the range 1--13~Gyr) for vast numbers of giant stars across the Galaxy.
We apply our model to $\sim$52,000 stars in APOGEE DR12, for which no direct mass and age information was previously available. We find that these estimates highlight the vertical age structure of the Milky Way disk, and that the relation of age with \aFe and metallicity is broadly consistent with established expectations based on detailed studies of the solar neighbourhood.
\end{abstract}


\section{Introduction} 
Obtaining accurate and precise ages for large numbers of stars in the Milky Way is a crucial ingredient in the comparison of observed data to galaxy formation simulations. It is also a first step towards understanding empirically how our Galaxy formed and how it evolved to its present-day structure. Stellar ages are unfortunately very hard to determine (see for example  \citealp{Soderblom2010}): they cannot be directly measured, and are always model-dependent.

A powerful way to measure ages for large samples of stars is to determine their location in the Hertzsprung--Russell diagram (HRD), and to compare this location with theoretical isochrones \citep{Edvardsson1993, Ng1998, Feltzing2001,Pont2004,Jorgensen2005,daSilva2006, Haywood2013, Bergemann2014}. This technique yields precise ages in regions of the HRD where isochrones of different ages are clearly separated, namely, at the main-sequence turn-off and on the subgiant branch. By contrast, on the red giant branch, isochrones of different ages are very close in temperature, so that they cannot be robustly used to determine ages. However, giant stars are crucial probes of the structure of the Milky Way, and routine age estimates for giants would be of enormous importance: their high luminosity makes them observable out to large distances; and the giants in old and young ($\sim 1$~Gyr) populations have comparable luminosities and colours, making their selection function far more age-uniform than in the case of turn-off stars. As a consequence, they are the primary targets in a growing number of surveys, including  the Apache Point Observatory Galactic Evolution Experiment (APOGEE), a high-resolution spectroscopic survey in the H-band \citep{Zasowski2013, Majewski2015}.

Because the mass of a star and its main-sequence lifetime are tightly correlated, ages for giants can be directly inferred from their mass. This has recently become the realm of asteroseismology, which can probe the internal structure of stars, not just  their surface properties. Thanks to the  CoRoT \citep{Baglin2006}, \textit{Kepler} \citep{Borucki2010}, and now K2 space missions, solar-like oscillations have been detected in thousands of red giants \citep[e.g.,][]{DeRidder2009, Hekker2009,  Bedding2010, Mosser2010,Hekker2011,Stello2013, Stello2015}, for stars up to 8 kpc from the Sun \citep{Miglio2013a}. Solar-like oscillations are pulsations that are stochastically excited by convective turbulence in the stellar envelope \citep[e.g.,][]{Goldreich1977, Samadi2001}. These oscillation modes are regularly spaced in frequency and contain information on the structure of the star. 

A first method to determine the properties of a star is to directly fit for the individual seismic frequencies \citep[e.g.,][]{Huber2013}, which gives a great precision on stellar masses and radii. However, it is very time-consuming and computationally-intensive and thus can only be done for small numbers of stars at a time. A simpler way to extract information from the power spectrum of the oscillations is to measure two global asteroseismic parameters: \dnu, the frequency separation of two modes of same spherical degree and consecutive radial order,  and \numax, the frequency of maximal oscillation power. A set of scaling relations directly links these two fundamental parameters to the mass and radius of a given star, so that the mass can be derived as
$
M \propto \nu_{\mathrm{max}}^3\ \Delta \nu^{-4} \ T_{\mathrm{eff}}^{1.5} 
$
(see Section 3.5 for more details).

Ages can then be inferred  by comparing the seismic data to theoretical isochrones  \citep[e.g.,][]{Stello2009,Kallinger2010,Basu2010, Quirion2010, Casagrande2014}, which leads to typical age uncertainties of the order of 30\%  \citep[e.g.,][]{Gai2011,Chaplin2014}.

Unfortunately, asteroseismology data are currently available only for relatively small samples of stars, located in a few different fields in the Milky Way. Future space missions like PLATO and TESS will have a larger sky coverage. In the meantime, it is very important to look for methods to determine stellar masses and ages that can be applied to large numbers of stars over a large volume of the Galaxy.

This paper is a first step towards using the information present in the APOGEE stellar spectra of giant stars to derive their masses and ages. Our work was inspired by \cite{Masseron2015}, who use the variations of carbon and nitrogen abundances between stars in the Milky Way's thin and thick disks to gather information on the relative ages of stars in both structures. Carbon and nitrogen are indeed expected to be good indicators of stellar masses: as a star arrives on the giant branch, its convective envelope extends deep into the star and brings up to the stellar surface material that has processed through the CNO cycle (this is called the first dredge-up). As a result of the convective mixing, the outer atmosphere will display signatures of this evolution, in particular a change in observed [C/N] ratio at the stellar surface \citep{Iben1965,Salaris2005}. Because the [C/N] ratio in the core and the depth reached by the dredge-up depend on stellar mass, the final [C/N] ratio at the surface depends on stellar mass. Since mass and age are closely related for stars on the giant branch, this also means that the [C/N] ratio can be used to infer stellar ages \citep{Salaris2015}.

There is however some scatter in model predictions, partly because the exact mixing processes affecting the surface abundances are still debated. It is thus tricky to directly use model predictions to link C and N abundances to stellar mass and age.
Our approach is to empirically determine the relationship between C and N abundances (and other stellar labels) and stellar mass in the APOKASC sample: there are currently 1,475 stars for which both APOGEE high-quality spectroscopic information and \textit{Kepler} asteroseismology information are available. In that sample, we find a strong correlation between mass, metallicity, and C and N abundances. The goal of this paper is to provide a fit to this relation, which can then be applied to a larger sample of APOGEE stars for which no \textit{Kepler} data is available. 

In a parallel paper, \cite{Ness2015b} use \textit{The Cannon} to confirm that APOGEE spectra contain information on stellar masses or ages. \textit{The Cannon} is a new  data-driven approach to determine stellar parameters from spectroscopic data \cite[see][]{Ness2015a}. With no prior knowledge of stellar evolution or stellar atmospheres, \textit{The Cannon} learns a mapping between wavelength and stellar parameters. \cite{Ness2015b} show that \textit{The Cannon} can also extract mass/age information from the APOGEE spectra, and that the spectral regions that contain the most mass information correspond to CN and CO molecules.

In this present work, our approach directly links stellar masses to the stellar parameters derived by the APOGEE pipeline from the stellar spectra, without using the spectra themselves. In Section \ref{sec:CNOcycles}, we review the theoretical expectations for the correlation between mass and [C/N] for giants. We then describe in Section \ref{sec:sample} the sample of stars we use, in particular, how we derive their masses and ages. In Section \ref{sec:observedCorrelation}, we present the observed correlations between mass and chemical abundances in the APOKASC sample. We then explain how we fit these correlations, discuss the performance of the models and the remaining biases (Section \ref{sec:model}). In Section \ref{sec:dr12}, we finally conclude the paper with an application of our models to the whole APOGEE sample, and present the correlations between the derived masses/ages with \aFe and metallicity, and with location in the Galaxy.

\section{CNO cycle, dredge-up and other mixing processes}
\label{sec:CNOcycles}
\subsection{The CNO cycle}
\begin{figure}
\centering 
\includegraphics[width=0.48\textwidth]{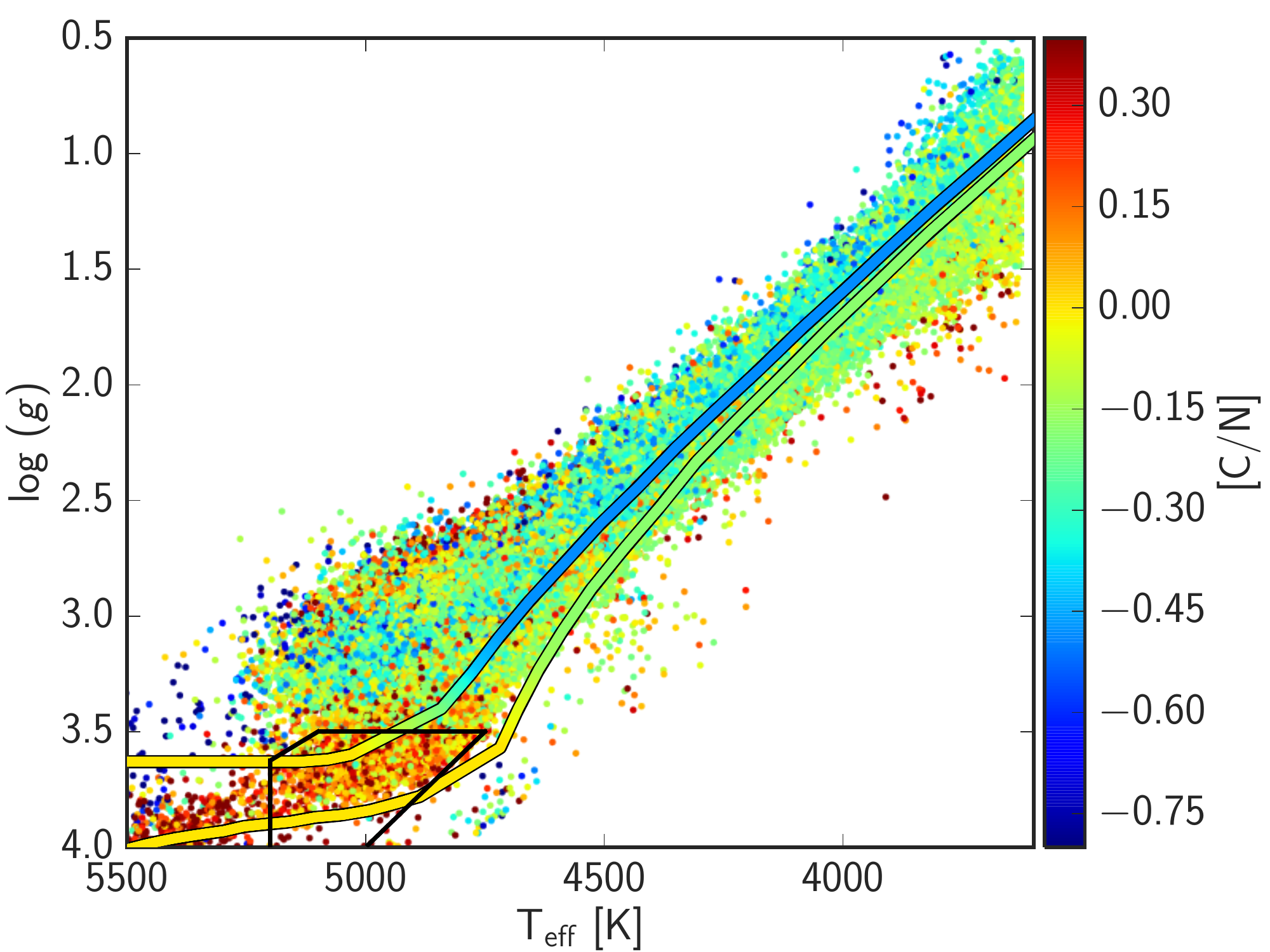}
\caption{H--R diagram for stars in APOGEE DR12, colour-coded by their surface [C/N] abundance. The two coloured lines show theoretical evolutionary tracks from \citet{Lagarde2012} for solar metallicity stars of 1 and 1.5 \msun, also colour-coded by the predicted surface [C/N].At the very bottom of the RGB, stars have a larger [C/N], which indicates that they have not experienced the first dredge-up yet. The black box shows how we select these pre dredge-up stars in DR12.}
\label{fig:selection}
\end{figure}

\begin{figure}
\centering 
\includegraphics[width=0.5\textwidth]{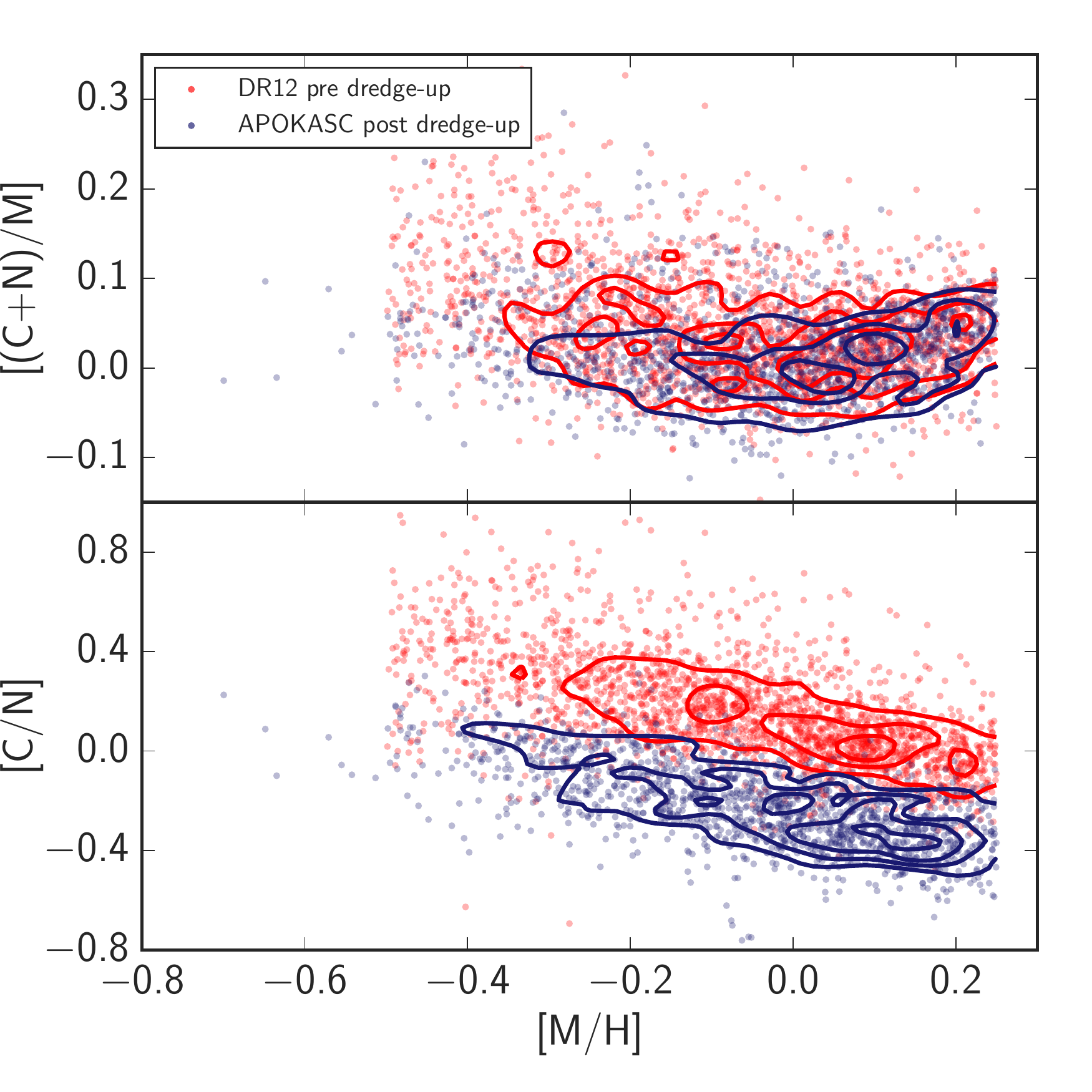}
\caption{Comparison of chemical abundances for giants in the APOKASC sample (giant stars that have experienced a dredge-up event, in blue points and blue contours) and for stars from APOGEE DR12 that have not experienced the dredge-up yet (in red, selected within the black box in Figure \ref{fig:selection}). The top panel shows that the two samples have a similar distribution of [(C+N)/M] as a function of metallicity [M/H]. The combined abundance of carbon and nitrogen does not change during the dredge-up, and hence reflects the birth properties of stars and how chemical evolution proceeds in the Milky Way. This shows that these two samples of stars have the same birth properties. The bottom panel shows that the APOKASC giants have a lower [C/N], which is due to the dredge-up that these stars have experienced.}
\label{fig:comp_nodredgeup}
\end{figure}

The CNO cycle consists in a series of nuclear reactions during which C, N and O atoms act as catalysts in the conversion of hydrogen to helium \citep[see for instance][]{Salaris2005}. While the total quantity of C, N, and O atoms is globally preserved during the nuclear reactions, their relative abundances evolve with time. More specifically, the slowest reaction in the CNO cycle corresponds to the proton capture on $^{14}$N, so that at equilibrium nitrogen becomes the most abundant element. In more detail, the CNO cycle produces an increase of the abundance in $^{14}$N in the stellar core, a decrease in $^{12}$C, a reduction of the ratio of $^{12}$C/$^{13}$C to $\sim$20--30 (to be compared to a solar value of $\sim$90, see \citealp{Asplund2009}), and a very slight change in $^{16}$O. 

At the end of the main sequence, the stellar interior is thus made of layers of material enriched in various elements. The exact shape of these layers can be affected by rotation during the main sequence (see for instance Figure 2 in \citealp{Charbonnel2010}).
The total amount of CNO-processed material in the core depends on stellar mass: stars with a higher mass have a larger central temperature, so that a larger fractional region of the stellar core reaches $^{12}$C burning temperatures. As a result, massive stars contain a higher fraction of nitrogen in their core.

\subsection{Post main sequence evolution}
As a star leaves the main sequence and starts to ascend the giant branch, its core contracts and the base of its convective envelope extends deeper into the star, to reach zones enriched in CNO-processed elements \citep{Iben1965}. This event, called the {\it first dredge-up}, results in a sharp change of surface abundances as the stellar surface becomes mixed with material enriched in nitrogen and depleted in carbon.

Figure \ref{fig:selection} illustrates the change of surface [C/N] for stars ascending the giant branch: we show a sample of stars from APOGEE DR12 in the log $g$ vs. \teff plane, colour-coded by their measured surface [C/N]. Stars at the very bottom of the RGB have a high [C/N] ratio, and this ratio quickly decreases for stars higher up on the RGB: the transition from one regime to the other corresponds to the first dredge-up. This figure also shows that the dredge-up happens within a similar range of log $g$ in the APOGEE data and in the stellar evolution models of \cite{Lagarde2012}.

Another way to visualize the effect of the first dredge-up is to compare [(C+N)/M] and [C/N] of stars before and after the dredge-up, as done in Figure \ref{fig:comp_nodredgeup}. Following the dredge-up, the surface abundance of [(C+N)/M] is unchanged because the total number of C, N, and O atoms is conserved (and the abundance of oxygen is only slightly affected by the dredge-up), but the ratio [C/N] clearly decreases. 

In canonical stellar evolution models, after the first dredge-up the surface abundances do not change any more until the AGB phase. However, observational data show that this is not the case: the carbon isotopic ratio and the abundance of carbon further decrease (and nitrogen increases) as stars climb the RGB 
\citep{Lambert1977, Suntzeff1981, Gilroy1989, Charbonnel1994, Gratton2000, Shetrone2003,Spite2006, Tautvaisiene2010,Angelou2012, Kirby2015}. These observations require non-canonical mixing mechanisms to move CNO-processed material from the hydrogen-burning shell into the convective envelope. Possible sources of deep mixing could be rotation  \citep{Charbonnel1995, Chaname2005} or thermohaline instabilities  \citep{Charbonnel2007}, although the importance of this process is debated \citep{Angelou2012}.

In any case, this additional mixing is only experienced by stars that go through an extended RGB evolution: this is the case of low mass stars (below $\sim$2--2.2 \msun). Indeed, at the end of the main sequence, these low mass stars have  an electron degenerate core and this core slowly grows in mass along the the RGB until it reaches a critical mass of 0.48~\msun, at what point helium burning is ignited. 
This event, the helium-core flash, marks the tip of the RGB. As mass loss increases rapidly as stars ascend the RGB, stars reaching the RGB-tip will experience the loss of a large fraction of their envelopes. After this, low-mass stars join the red clump.

Stars that are more massive than $\sim$2--2.2 \msun only go through the first dredge-up without any further mixing processes or mass loss because their RGB evolution is very short. Indeed, these stars are massive enough to have a non-degenerate core and to ignite helium gently; once this is done they populate the secondary red clump \citep{Girardi1999}.

\subsection{Correlation between stellar mass and surface abundance of C and N}

\begin{figure}
\centering 
\includegraphics[width=0.5\textwidth]{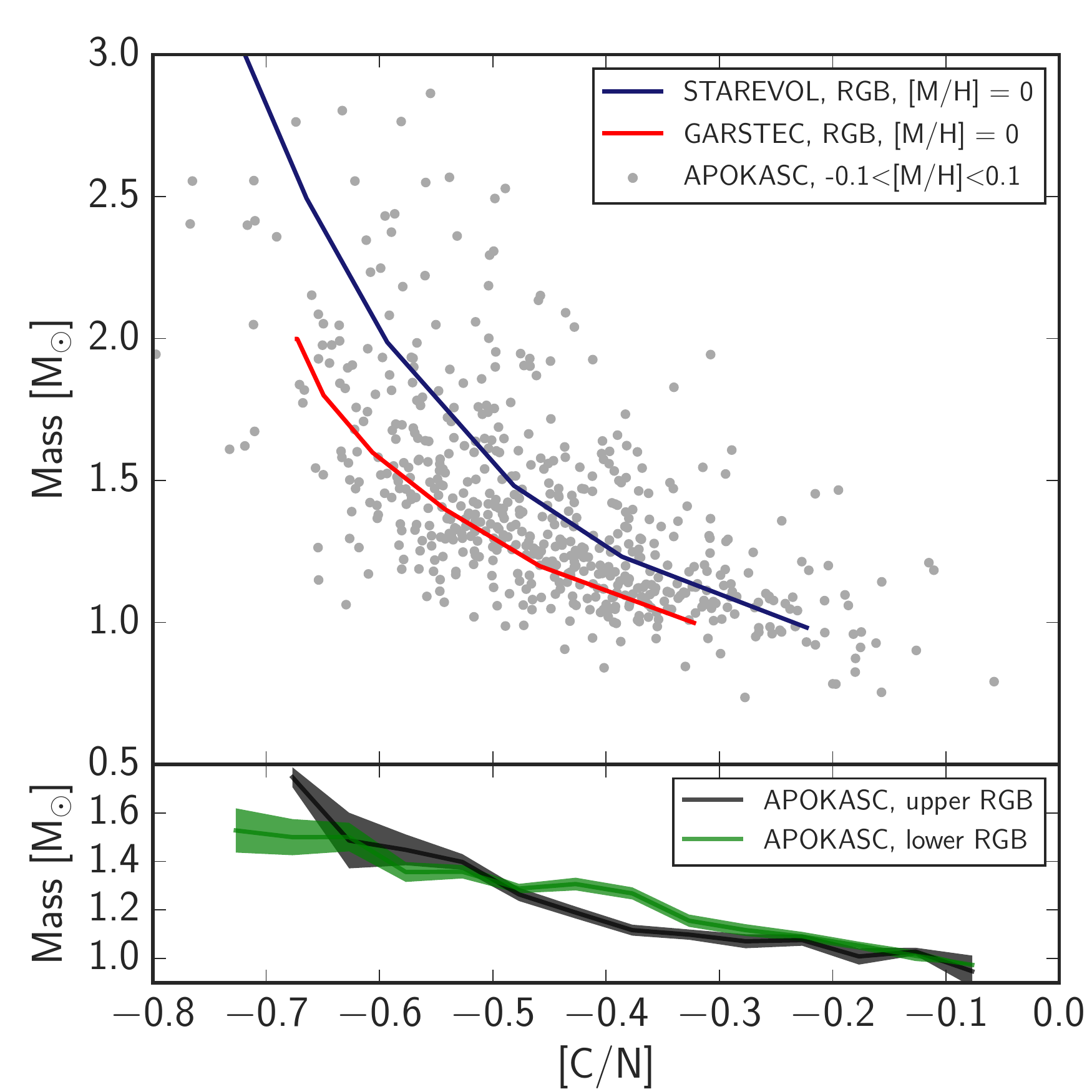}
\caption{Relationship between stellar mass and surface [C/N] in the APOKASC sample. The top panel compares the APOKASC stars (grey dots, here limited to $-0.1<[M/H]<0.1$) to stellar evolution models of \citet{Weiss2008} in red  and \citet{Lagarde2012} in blue (we show the "standard" models only including the first dredge-up). For the data and the models to match, the observed [C/N] has to be increased by 0.2: this reflects calibration issues for C and N abundances in APOGEE DR12. Models and data all show a decrease of [C/N] with increasing stellar mass, although models do not agree on the magnitude of the predicted decrease. The bottom panel compares the mean mass as a function of [C/N] for APOKASC stars on the upper and lower RGB: in the presence of extra-mixing processes along the RGB, stars on the upper RGB would be expected to have a lower [C/N] at fixed stellar mass, but there is no evidence for this in the current data.}
\label{fig:dredgeup}
\end{figure}

The surface abundances of a star after the first dredge-up depend both on the distribution of CNO-processed material within the core at the end of the main sequence, and on the depth reached by the base of the convective envelope during the dredge-up. Both of these aspects depend on the mass of the star: stars of increasing mass contain a higher fraction of nitrogen in their core, and have a convective zone that extends much deeper. Metallicity, helium fraction and abundance in $\alpha$ elements also influence the depth reached by the envelope because they impact its opacity \citep{Sweigart1989,Boothroyd1999}, but these are only minor effects for stars below 3 \msun \citep{Charbonnel1994,Karakas2014}.

As a result, after the first dredge-up, the surface of higher mass stars is comparatively richer in N and poorer in C with respect to lower mass stars.  As an example, we show in Figure \ref{fig:dredgeup} the relationship between mass and the [C/N] ratio after the first dredge-up in the models of \cite{Lagarde2012} as well as models computed with the GARching STellar Evolution Code (GARSTEC, \citealt{Weiss2008})  for stars of solar metallicity: there is a decrease of [C/N] with increasing stellar mass. This Figure also compares the models with observed mass and [C/N] from our APOKASC sample (see Section \ref{sec:sample} for explanations of how these quantities are derived): data and model predictions are roughly in agreement. There are however  variations between models. This, together with potential calibration issues of abundances in APOGEE (in Figure \ref{fig:dredgeup}, we have to shift the observed [C/N] by 0.2 dex to match model predictions),  makes it difficult to directly use model predictions to translate an observed [C/N] into mass or into age, as suggested by \cite{Salaris2015}.

Another potential hurdle in the use of [C/N] to determine stellar masses is that the relation between abundances and mass might depend on stellar evolutionary phase, as we discussed in the previous section. Stars in the upper RGB would both undergo extra mixing and mass loss compared to stars on the lower RGB. Stars in the red clump would have the largest mass loss, while stars in the secondary clump would have no extra mixing and no mass loss, and should be similar to stars on the lower RGB. We do not have a sample of massive stars on the lower RGB to compare to our secondary clump stars, but we can compare the mass and surface [C/N] of stars on the lower and upper RGB (lower panel in Figure \ref{fig:dredgeup}). In the APOKASC sample, we find no significant evidence for a different relation between mass and [C/N] in the upper and lower RGB. This is slightly unexpected, but could reflect the lower sensitivity to extra mixing of [C/N] compared to $^{12}$C/$^{13}$C \citep[e.g., ][]{Tautvaisiene2010}, and the inefficiency of extra mixing processes at the relatively high metallicities of our sample \citep{Gilroy1989, Gratton2000, Charbonnel2007, Martell2008}.

Because of these uncertainties, in this paper, we decide not to rely on theoretical models to connect the masses of giant stars to the abundance of carbon and nitrogen at their surface. Instead, we explore this correlation empirically using the APOKASC sample.

\section{The APOKASC sample}
\label{sec:sample}
The APOKASC project is the spectroscopic follow-up by APOGEE (\citealt{Majewski2015}, as part of the third phase of the Sloan Digital Sky Survey, SDSS-III; \citealp{Eisenstein2011}) of stars with asteroseismology data from the \textit{Kepler} Asteroseismic Science Consortium (KASC).  The first version of the APOKASC catalogue \citep{Pinsonneault2014} contains seismic and spectroscopic measurements for 1,989 giants, with the spectroscopic information corresponding to APOGEE's Data Release 10 (DR10; \citealp{Ahn2014}). In this present work, we keep the same original sample of 1,989 stars and their seismic parameters, but update their \teff and abundances to DR12 values \citep{Alam2015,Holtzman2015}. This follows the same procedure as in \cite{Martig2015}.

\subsection{Seismic parameters from \textit{Kepler}}\label{sec:keplerdata}
The 1,989 giants have been observed by \textit{Kepler} over a total of 34 months (Q0--Q12) in long cadence mode, i.e., with a 30 minute interval \citep[e.g., ][]{Jenkins2010}. The raw light curves were prepared as described in \cite{Garcia2011}, and the seismic parameters \numax and \dnu   were then measured using five different techniques \citep{Huber2009,Hekker2010, Kallinger2010, Mathur2010,Mosser2011}. The final values of \numax and \dnu given in the APOKASC catalogue correspond to the ones obtained with the OCT method from \cite{Hekker2010}. The other four techniques are only used in an outlier rejection process (stars with \numax values that differ significantly from one technique to another are removed from the sample) and to estimate systematic uncertainties on the measured parameters.

\subsection{Spectroscopic parameters from APOGEE}\label{sec:apogeedata}
APOGEE is a high-resolution ($R= 22500$) \textit{H}-band stellar survey which uses a multi-fiber spectrograph attached to the 2.5\,m SDSS telescope \citep{Gunn2006}. The raw spectra are first processed by the APOGEE data reduction pipeline, as described in \cite{Nidever2015}. Stellar parameters are then derived with the APOGEE Stellar Parameter and Chemical Abundances Pipeline (ASPCAP; \citealp{Meszaros2013}, Garc\'{\i}a P\'{e}rez et al., in preparation). 
ASPCAP compares the observed spectra to a large grid of synthetic spectra \citep{Meszaros2012,Zamora2015} to determine the associated main stellar parameters. This synthetic grid has six dimensions: \teff, log $g$, metallicity [M/H],	as well as enhancement in \al-elements \aFe, in carbon [C/M] and in nitrogen [N/M]. A $\chi^2$ optimization finds the best fit spectrum, and the corresponding stellar parameters are assigned to the observed star. 

In addition to these parameters, DR12 also provides calibrated abundances for some elements as well as post-calibrated values of \teff and log $g$ by using literature studies of well-known star clusters, and the APOKASC catalogue as reference for log $g$. In this present work however, we always use the {\it raw} values to ensure that they are all self-consistent. In practice, this means that we use the values from the FPARAM array in DR12\footnote{See http://www.sdss.org/dr12/irspec/parameters/}.

Finally, the ASPCAP pipeline also returns uncertainties on stellar parameters and abundances. It seems however that the formal errors from the 6D fits to the spectra underestimate the true uncertainties on the stellar parameters: globular and open clusters are expected to be chemically homogeneous but the spread in chemical abundances for stars within a given cluster is larger than the formal errors \citep{Holtzman2015}. A corrected (empirical) estimate of the uncertainties is then provided by measuring the spread of abundances within star clusters. Unfortunately, this procedure does not work for [C/M] and [N/M] because giant stars in clusters are expected to have an intrinsic spread in [C/M] and [N/M]. Thus, while DR12 provides uncertainties for [C/M] and [N/M] (with mean values of 0.04 and 0.07 dex respectively), these values are probably underestimated. A further analysis by \cite{Masseron2015} shows that the precision on [C/N] is still probably better than 0.1 dex.

\subsection{Carbon and nitrogen abundances}\label{sec:CNabundances}
The abundances of carbon and nitrogen are mainly measured from molecular lines of CO and CN. Because these lines become very weak for hot stars at low metallicity,
the minimum abundance of [C/M] that can be measured depends on \teff and [M/H] \citep[see][for a discussion of this issue]{Meszaros2015}. We want to eliminate from our sample stars with only an upper limit measurement on [C/M] (and a lower limit on [N/M]), as such measurements may introduce a bias in our analysis if left in the sample. Thus, following \cite{Meszaros2015} we remove from our sample selection stars with a raw \teff greater than 4550 K if -1$<$[M/H] $<$-0.5.

For stars with [M/H] $>$-0.5, we performed our own upper limit tests by selecting stars that have \teff greater than 4550 K and [C/M]$<$-0.1. We performed our own $\chi^2$ search using Autosynth \citep{Meszaros2015} to fit individual CO lines in the APOGEE windows from [C/M]=-0.4 to +0.7. By inspecting the $\chi^2$ as a function of [C/M] we found that the minimum [C/M] possible to measure is on the level of -0.4 to -0.5~dex, far smaller than the most carbon poor star found in our sample. Thus, it was determined that no temperature cut is necessary near solar metallicity and below 5000~K.

Even for stars with ``good'' measurements, there is a zero-point issue with the absolute [N/M] values: they are $\sim 0.2$ dex too low compared to literature values as discussed by \cite{Holtzman2015} and \cite{Masseron2015}. This systematic offset does not impact this study. However, such offsets also mean the results of our fits cannot be blindly applied to another survey, or even to another data release of APOGEE: the abundances would first need to be put on the same scale, for instance using \textit{The Cannon} \citep{Ness2015a}.

\subsection{Sample selection}\label{sec:samplecuts}
Our goal is to learn an empirical relation between C, N abundances and stellar parameters. Therefore, we need a reliable sample of stars.
Starting from the APOKASC--DR12 giant stars sample, we first eliminate stars for which any of the ASPCAP flags are set to WARNING or BAD (this signals potential problems with the determination of spectroscopic parameters), as well as stars for which the spectra have a signal-to-noise ratio below 100. We also tried other quality cuts using the $\chi^2$ value of the ASPCAP best fit to the spectra, and using the number of times a given star was observed, but none of those impacted our results.

To ensure the good quality of the seismic masses we derive, we remove stars with relative uncertainties on \dnu and \numax greater than 5\%, and the most metal poor stars ([M/H] $< -1$), for which the standard seismic scaling relations might be less accurate \citep{Epstein2014}. Finally, we  exclude the fast rotating stars  (14 rapid and 12 additional anomalous rotators) identified by \cite{Tayar2015}. Such stars might be accreting mass from a companion, so that their surface properties might not correspond to their mass and evolutionary stage.  Out of the 1,989 stars with seismic and spectroscopic information, 1,475 stars remain; these objects form the sample used in this paper. 

\subsection{Determining masses from seismic scaling relations}\label{sec:scalingRelations}

\begin{figure}
\centering 
\includegraphics[width=0.5\textwidth]{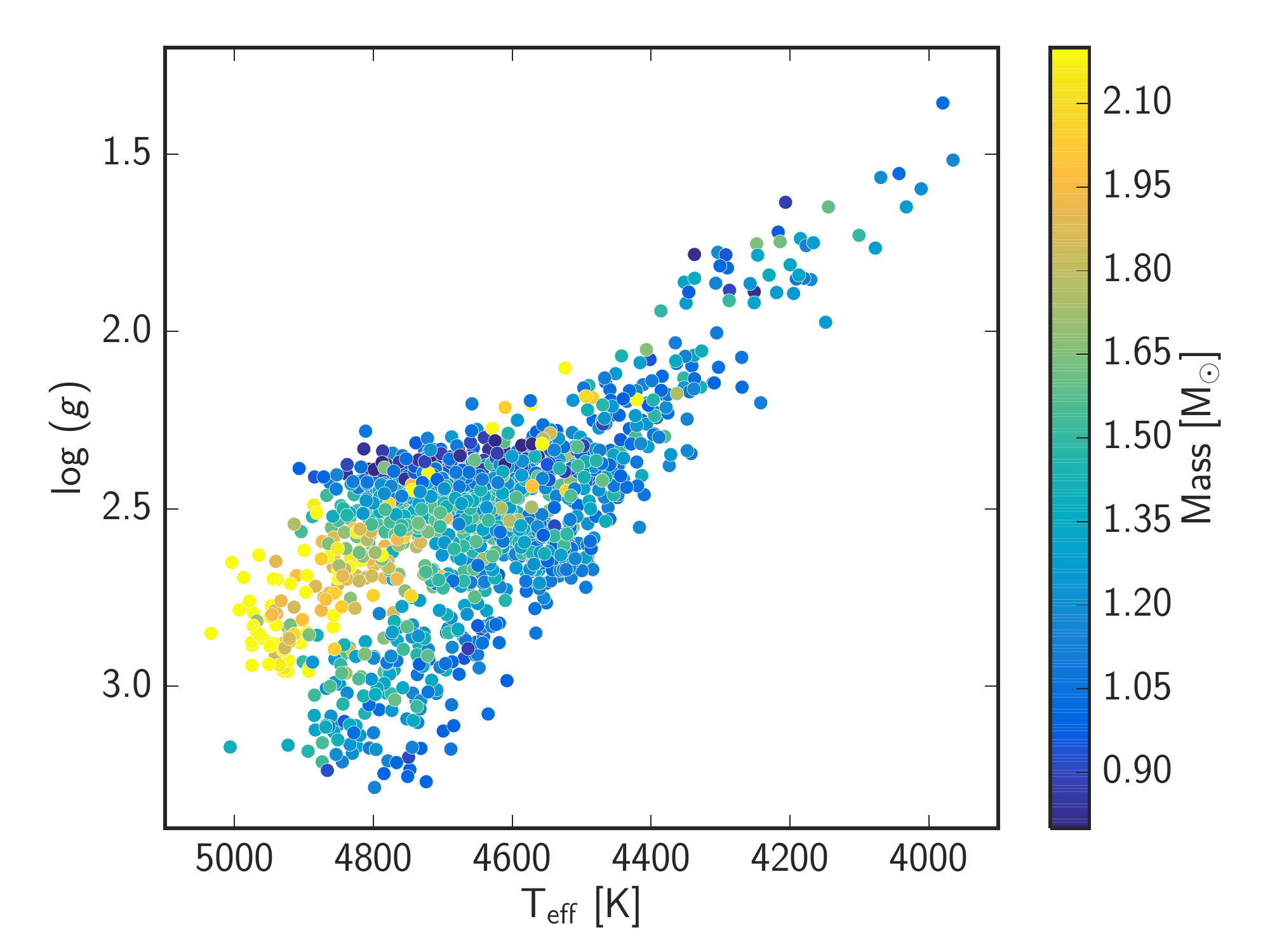}
\caption{Surface gravity as a function of effective temperature for stars in the APOKASC sample. The surface gravities are determined from the \textit{Kepler} seismic parameters, while \teff is derived from from the APOGEE spectra. The points are colour-coded by mass.}
\label{fig:HRmass}
\end{figure}
Solar-like oscillations can be described by two main global asteroseismic parameters, \dnu and \numax. 
The large frequency separation, \dnu, depends on the sound travel time from the centre of the star to the surface, and is thus related to the stellar mean density: \citep{Tassoul1980,Ulrich1986,Kjeldsen1995},
\begin{equation}\label{eq:dnu}
\Delta \nu \propto \rho^{1/2} \propto M^{1/2} R^{-3/2}\ .
\end{equation}
On the other hand, \numax (the frequency of maximal oscillation power) is related to the acoustic cut-off frequency \citep{Brown1991}, which mainly depends on surface gravity and temperature \citep{Kjeldsen1995,Belkacem2011}:
\begin{equation} \label{eq:numax}
\nu_{\mathrm{max}} \propto g T_{\mathrm{eff}}^{-1/2} \propto  M R^{-2}T_{\mathrm{eff}}^{-1/2}\ .
\end{equation}

The standard seismic scaling relations, Equations 1 and 2, can be combined to derive the mass of a star as:
\begin{equation} \label{eq:mass}
M= \left( \frac{\nu_{\mathrm{max}}}{\nu_{\mathrm{max,\odot}}}\right)^3\  \left( \frac{\Delta \nu}{\Delta \nu_{\odot}}\right)^{-4} \ \left( \frac{T_{\mathrm{eff}}}{T_{\mathrm{eff,\odot}}}\right)^{1.5} \ .
\end{equation}
We adopt  $T_{\mathrm{eff,\odot}}=5777$ K, $\nu_{\mathrm{max,\odot}}=3140\ \mu$Hz, $\Delta \nu_{\odot}=135.03\ \mu$Hz. The solar values  $\Delta \nu_{\odot}$ and $\nu_{\mathrm{max,\odot}}$ are the ones used to build the APOKASC catalogue and were obtained by \cite{Hekker2013} with the OCT method. As an exception to the rule generally used in this paper, we do not use here the raw ASPCAP values of \teff, but use instead the values that are calibrated to match the photometric temperatures calculated from the 2MASS  J-K$_s$ colour \citep[as in][]{Gonzalez2009}. This ensures that \teff is closer to the ``true'' physical scale.
We derive the mass uncertainty from the uncertainties on \numax, \dnu, and \teff, which have average values of 3.1\%, 2.4\%  and 1.9\%, respectively; this leads to an average mass uncertainty of 0.2 \msun (or 14\%). 

While scaling relations have been widely used to determine stellar masses, small deviations to the \dnu scaling relation have been proposed, both based on studies of stellar models and on the determination of masses for stars in open clusters. \cite{White2011} use stellar models to show that the relation between \dnu and stellar density matches the standard relation for solar type stars on the main sequence, but that deviations of the order of 2\% in the relation between \dnu and $\sqrt{\rho}$ exist for stars on the giant branch. This translates into a mass 8\% smaller than predicted by the scaling relations. \cite{Huber2013} find a similar offset  when comparing the mass of the red giant Kepler-56 obtained from the scaling relations to the mass obtained from an analysis of the individual seismic frequencies. 

It seems also that deviations to the standard relations are different for red clump (RC) and RGB stars. This is not unexpected, because these two types of stars have very different internal structures, hence different temperature and sound speed profiles. \cite{Miglio2012} studied the mass of stars in open clusters and found an offset between the mass of RC and RGB stars that cannot be explained by mass loss alone. Further models by \cite{Miglio2013b} also suggest  a different offset in the \dnu scaling relation for RC and RGB stars (in the sense that RC masses are underestimated and RGB masses are overestimated by the standard relation).

To apply modifications to the scaling relation to our sample of stars, we first need to identify RC and RGB stars. While some of the stars in the APOKASC catalogue have such a label ('CLUMP' or 'RGB' in the catalogue), it is not the case for all stars. We classify stars as RC stars if they are identified as such by their seismic properties, or if they were identified as being in the RC region of the H--R diagram by \cite{Bovy2014} \footnote{For this, we simply check which stars of the APOKASC sample are also in the RC catalogue from \cite{Bovy2014}}, or  if $\log g <0.00221 \times T_{\mathrm{eff}}-7.85$. All other stars are identified as RGB stars.

For all stars identified as RGBs following these criteria, we reduce the mass by 8\%, while we leave the mass of RC stars as predicted by the scaling relations. 
 Figure \ref{fig:HRmass} shows the distribution of our sample in the surface gravity versus \teff plane, with stars colour-coded by their mass. Stars in the red clump show a correlation between mass and log $g$, with the most massive stars being located in the secondary clump at log $g$ slightly below 3.

\subsection{From mass to age}

In principle, ages of giant stars can be derived by fitting isochrones to the location of stars in the HR diagram. However, as discussed in \cite{Martig2015}, the location of the RGB is quite uncertain in stellar evolution models, and uncertainties on the measurements of \teff exacerbate the problem. We thus choose a simple way to translate mass into age based on the stars' age as a function of stellar mass for different phases of stellar evolution (at the bottom of the RGB, in the RC, and at the tip of the asymptotic giant branch --- AGB).

For a given metallicity, we use the relation between mass and age as given by the PARSEC isochrones\footnote{http://stev.oapd.inaf.it/cmd} \citep{Bressan2012} using a mass loss parameter $\eta=0.2$. Although the mass loss parameter is an uncertain quantity, such a relatively low value is favoured by the study of \cite{Miglio2012}. We use a set of isochrones regularly spaced by log(age/yr)=0.015 from 100 Myr to 13 Gyr, and ranging from [M/H]=-1 to 0.5 in bins of 0.1. For each star, we use the set of isochrones closest to its metallicity (without interpolating between sets of different metallicity).

For RC stars (as defined in the previous section), we use the relation between mass and age in the RC. Ten stars have a mass too small to be consistent with the isochrones we use: either their measured stellar mass is too low (from measurement errors, or because the scaling relations should be modified also for the RC stars), or they have lost more mass than prescribed by our chosen set of isochrones. We attribute an age of 13 Gyr to these stars.

For the rest of the stars, we use the relation between mass and age at the bottom of the RGB. 32 stars have a too low mass to be consistent with isochrones. For the stars with log(g)$<2$ we use the relation between mass and age at the tip of the AGB, for stars with log(g) between 2 and 2.7, we use the relation for RC stars. Stars that cannot be attributed an age in any of these ways are given an age of 13 Gyr.

For each star we estimate an age uncertainty by translating  into age the upper and lower limit of our mass uncertainty range following the procedure we have just described.

\begin{figure*}
\centering 
\includegraphics[scale=0.35]{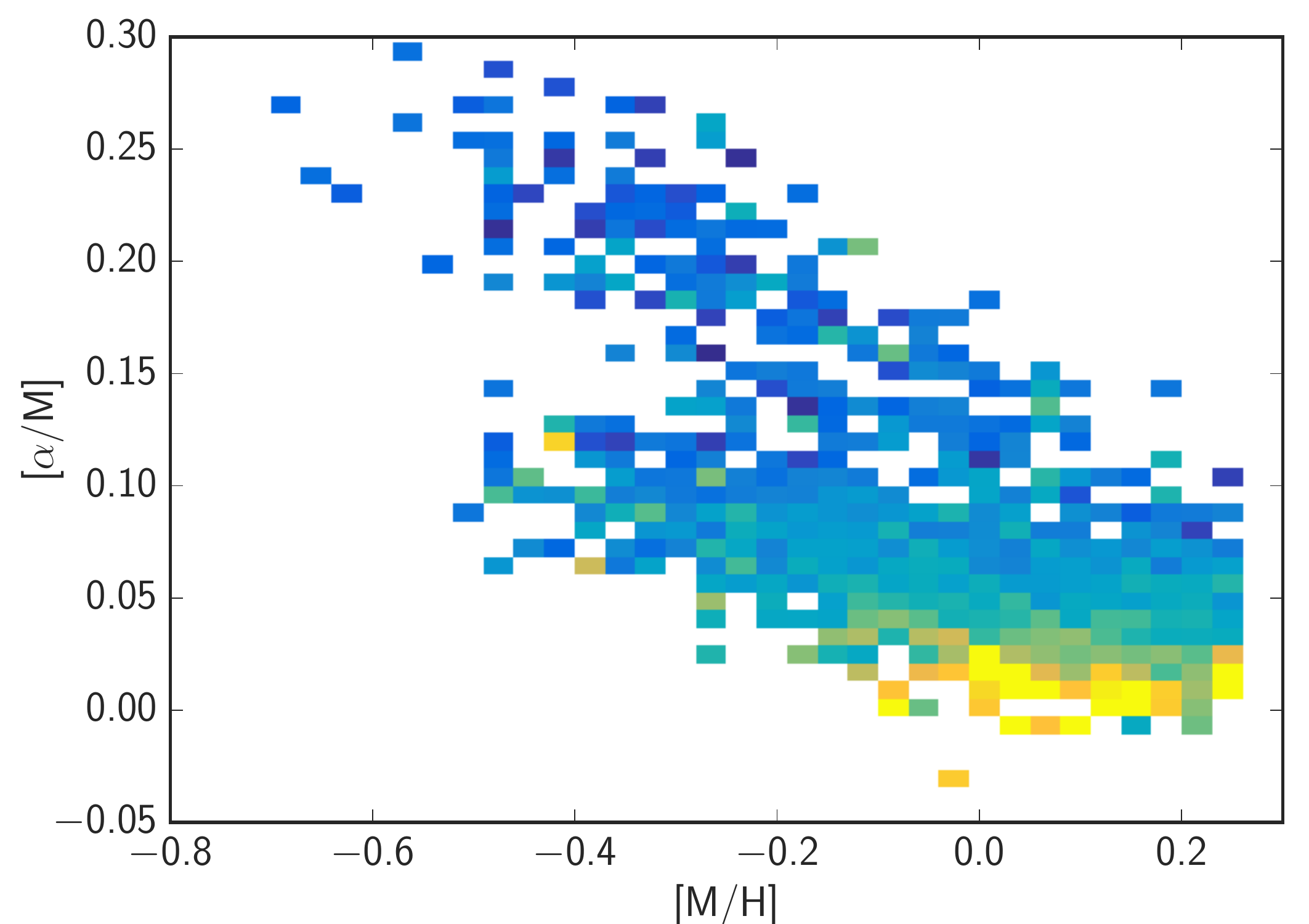}
\includegraphics[scale=0.35]{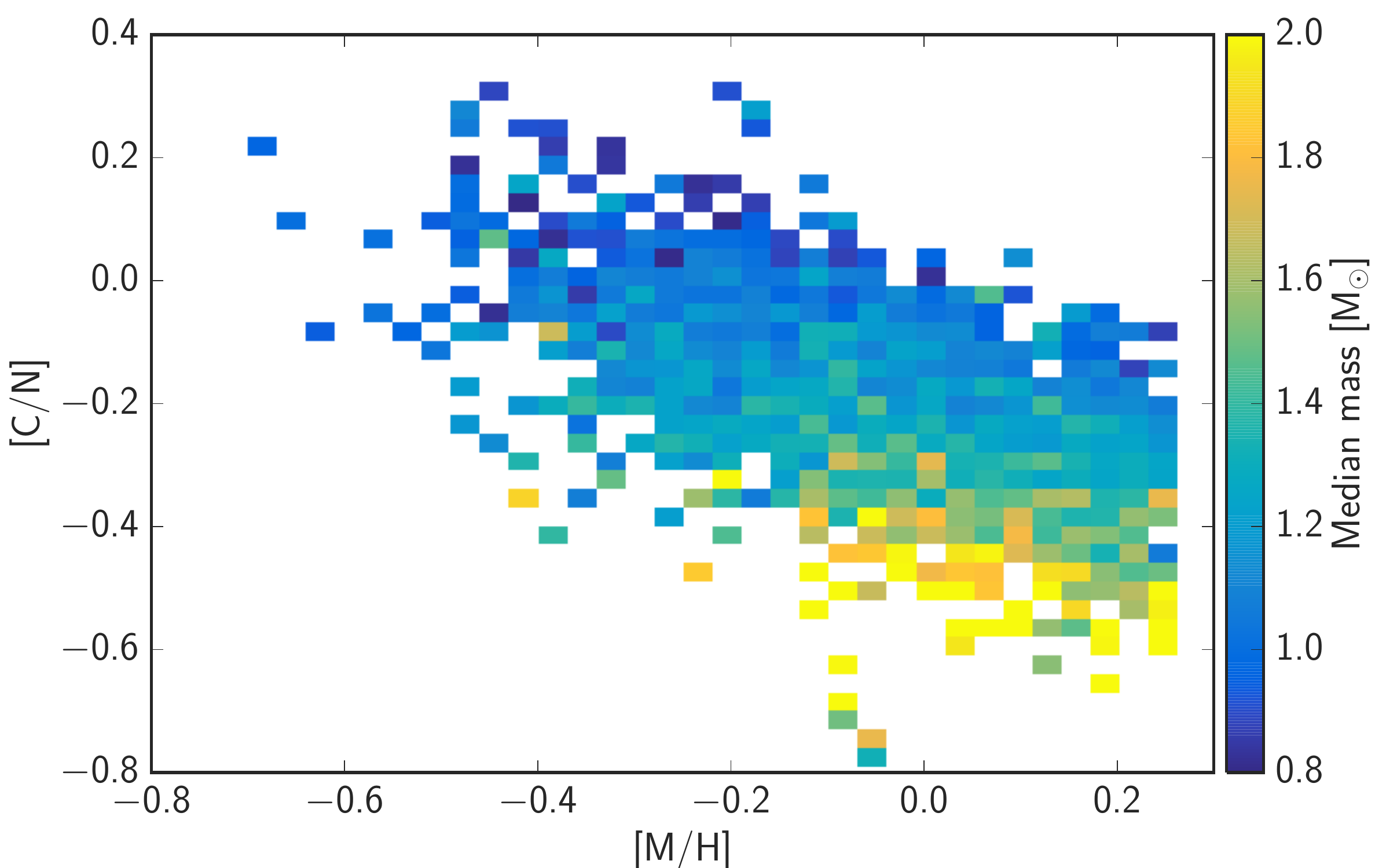}

\includegraphics[scale=0.35]{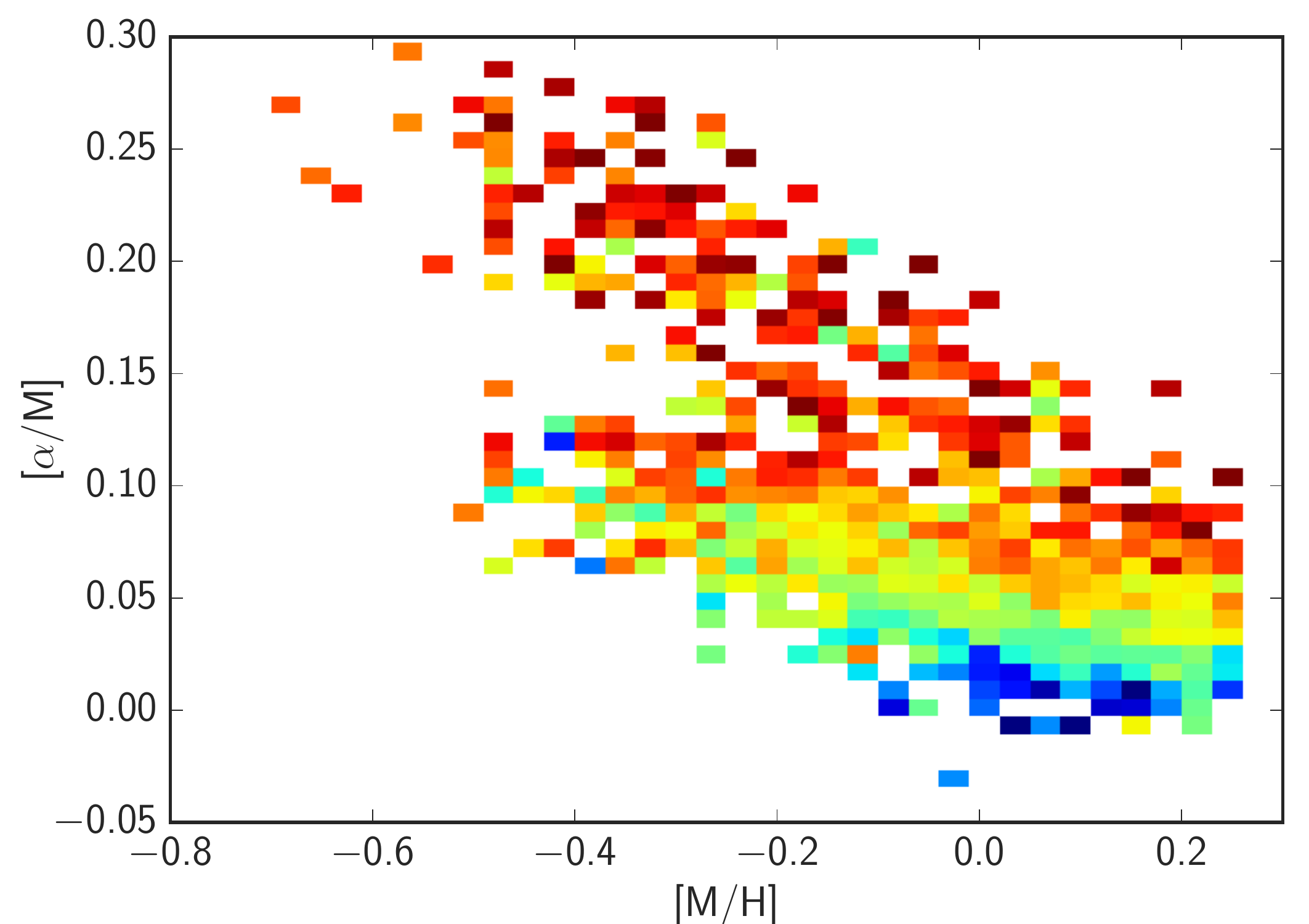}
\includegraphics[scale=0.35]{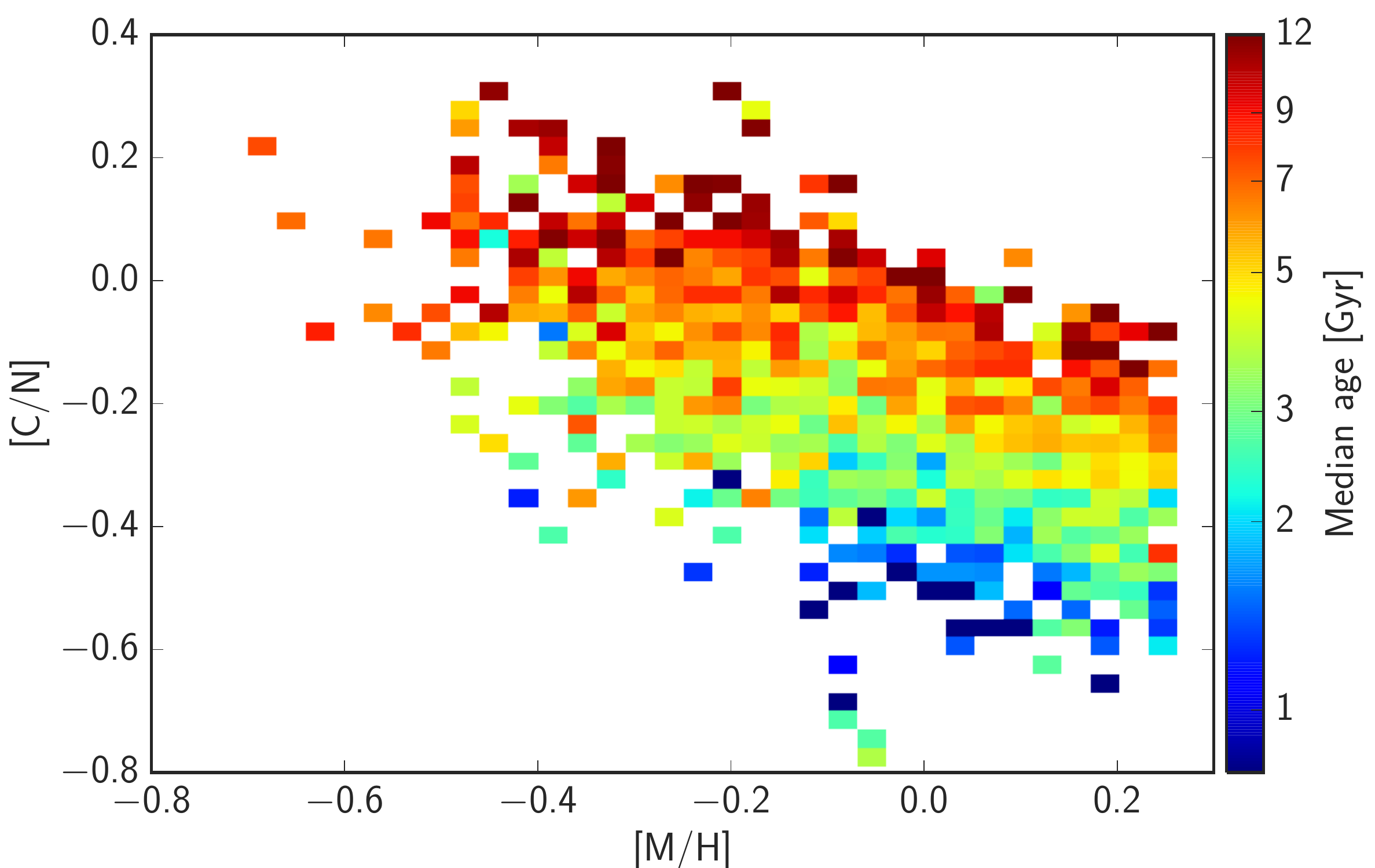}

\caption{Relationship between surface abundances and direct asteroseismic mass (top row) or age (bottom row), shown for 1,475 of stars in the APOKASC sample. The left column shows \aFe as a function of [M/H] while the right column shows [C/N] as a function of [M/H]. While age and \aFe are correlated because of Galactic chemical evolution, the correlation between [C/N] and age is due to internal stellar evolution. As expected from stellar models, stars with a high [C/N] have a small mass and a large age. In a first approximation, stars of a given age are found along parallel diagonal lines in this plane. The upper edge of the stellar distribution is then determined by the age of the Universe (and the smallest mass a star can have and still reach the giant branch in $\sim$13 Gyr).}
\label{fig:alpha_mass}
\end{figure*}

\begin{figure*}
\centering 
\includegraphics[width=0.48\textwidth]{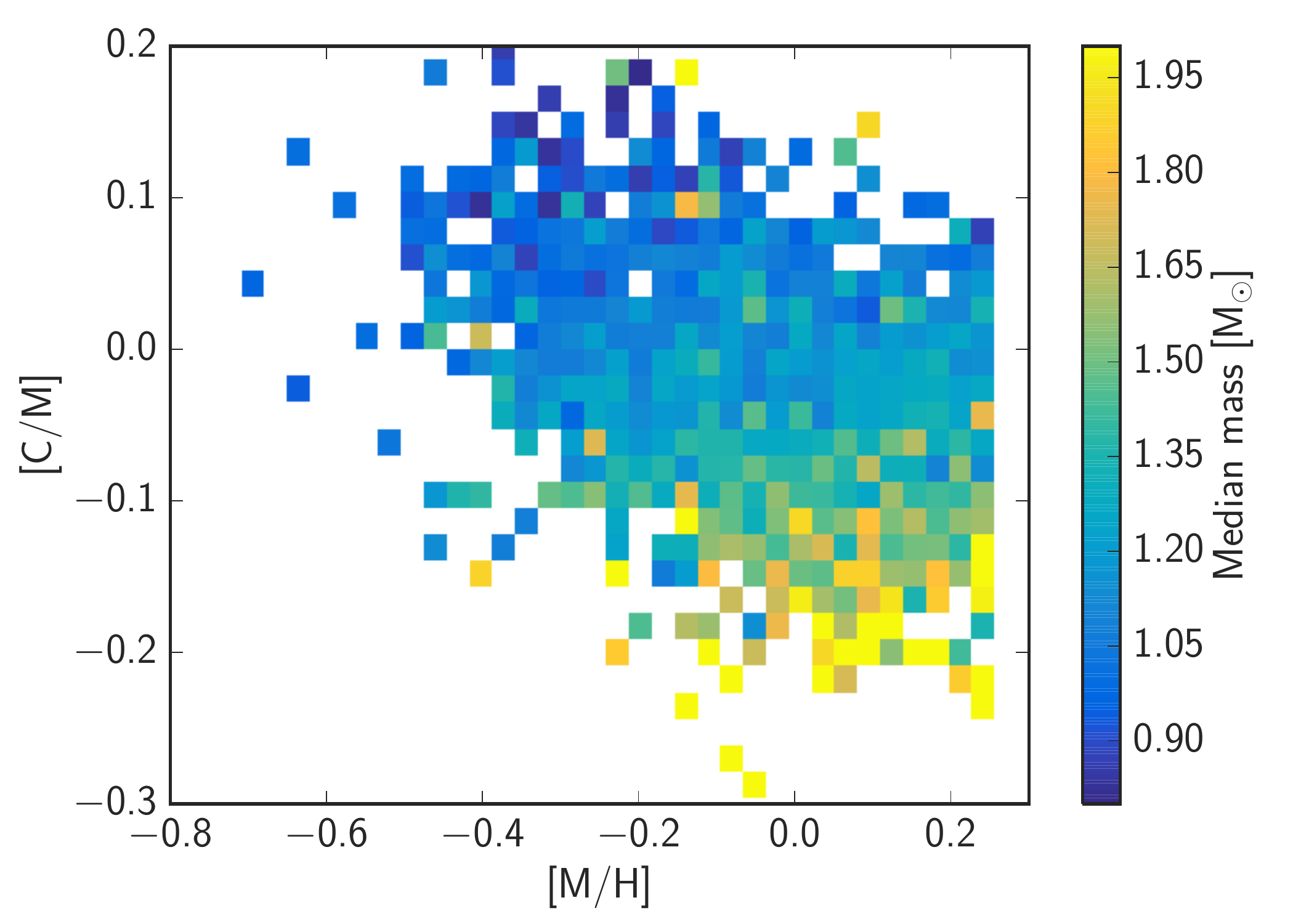}
\includegraphics[width=0.48\textwidth]{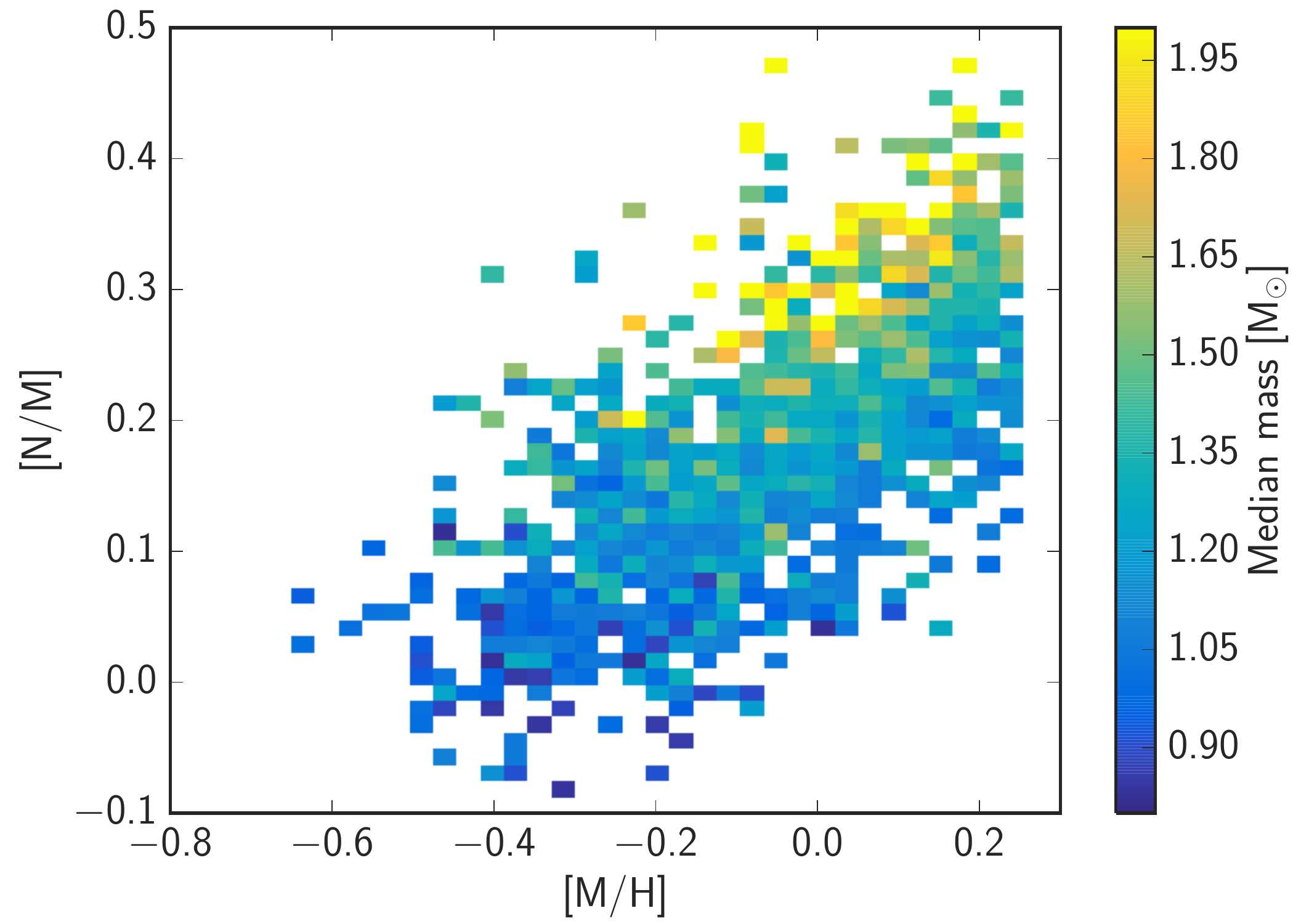}
\caption{Stellar mass in the [C/M]-[M/H] plane (left panel) and [N/M]-[M/H] plane (right panel). This shows that both carbon and nitrogen abundances contain mass information.}
\label{fig:FeH_CM_NM}
\end{figure*}
\section{An observed correlation between mass and chemical abundances}
\label{sec:observedCorrelation}

As a sanity check, we show in Figure \ref{fig:alpha_mass} the relation between \aFe, [M/H] and mass (left column, top row) or age (left column, bottom row) for the 1475 giants we have selected from  the APOKASC sample. This Figure shows that, as expected  from previous studies and from chemical evolution models, the \al-rich sequence mostly contains low-mass, old stars. As \aFe decreases, stars become more massive and younger. To investigate the scatter in age within a bin of \aFe and [M/H], we first group bins together using a Voronoi binning algorithm \citep{Cappellari2003} so that each new bin now contains 8 stars on average. Within each of the new Voronoi bins, we then compute the median mass and age for all stars in that bin.   If we compare the values of mass and age for each star to the median mass and age of stars in the same Voronoi bin, we find a  median scatter in mass of 9 per cent, and a median age scatter of  26 per cent.

As we mentioned in Section \ref{sec:CNOcycles}, stellar evolution models predict a correlation between mass, carbon and nitrogen abundances. This correlation arises from internal evolution of the stars, hence from a different origin than the \aFe and mass correlation. Indeed, the latter does not reflect the stellar evolution but the composition of the material from which stars are born.

In the right column of Figure \ref{fig:alpha_mass}, we show the relation between [C/N], [M/H] and mass or age for the APOKASC giants.  For a given metallicity, a low [C/N] ratio corresponds to a high stellar mass and a small age. Similarly as in Figure \ref{fig:dredgeup}, the magnitude of the decrease of [C/N] with stellar mass is consistent with stellar evolution models.
 
We build Voronoi bins in the same way as in the \aFe--[M/H] plane, and then compare the values of mass and age for each star to the median mass and age of stars in the same Voronoi bin. The  median scatter in mass is  9 per cent, and the median age scatter is  25 per cent. This is similar to  the age and mass scatter in the \aFe--[M/H] plane. 

Figure \ref{fig:FeH_CM_NM} demonstrates that the correlation between mass and [C/N] arises both from a decrease of [C/M] with mass and an increase of [N/M] with mass: both elements contain mass information, as predicted by stellar evolution models.

\section{Modelling the correlation between mass and abundances of carbon and nitrogen}
\label{sec:model}

\subsection{Fitting procedure: a polynomial feature regression}

We want to generate a model that can predict masses and/or ages from a set of spectroscopic observables. We start with only considering element abundances, namely  [M/H], [C/M], and [N/M]. The set of abundances could be larger, but in this present work, we want to stay close to the stellar evolution physics detailed above. Therefore, we explore the construction of such predictive model with a minimum number of chemical elements. 

Our model is relatively simple. We define it from a polynomial combination of the different features (e.g., chemical elements), which also includes cross-terms between the different dimensions. This allows us to effectively expand our dataset to non-linear combinations of our initial dimensions.

More specifically, we denote the coefficient in front of the $i$-th label $l_i$ as $k_i$. Then, the predicted value $y$ (i.e., age or mass) is given by:
$$y = \sum_i k_i\,l_i + \epsilon$$
which we can write with vectors as $$y = \vec{K}\cdot\vec{L} + \epsilon$$ 

This corresponds to a simple linear regression (linear in the parameters, $\vec{K}$), in which $\epsilon$ is a constant allowing us to account for a non-zero offset in this relation. The training data is provided in Table \ref{tab:data}.

We estimate the internal uncertainties on the fit parameters and on the predicted values by drawing 100 fiducial samples from the data (assuming Gaussian errors on both the input labels and on masses or ages), and performing a set of 100 linear regressions, giving 100 different realizations of both the model and the predicted mass or age. We use the standard deviation of these 100 different predicted masses for each star as an estimator of the mass internal uncertainty in the model.

We also validate our model through cross-validation. This is a way to test how well our model would apply to other datasets, and to give a better estimate of the model performance and external errors. We use a Leave-One-Out Cross-Validation (LOOCV) algorithm: for a set of $N$ stars, this consists of training the model on $N-1$ stars, and testing the performance on the last star, i.e., measuring the error the model makes when predicting parameters of that particular star. This step is repeated $N$ times, once per star from the training dataset. 

\begin{figure*}
\centering 
\includegraphics[width=0.48\textwidth]{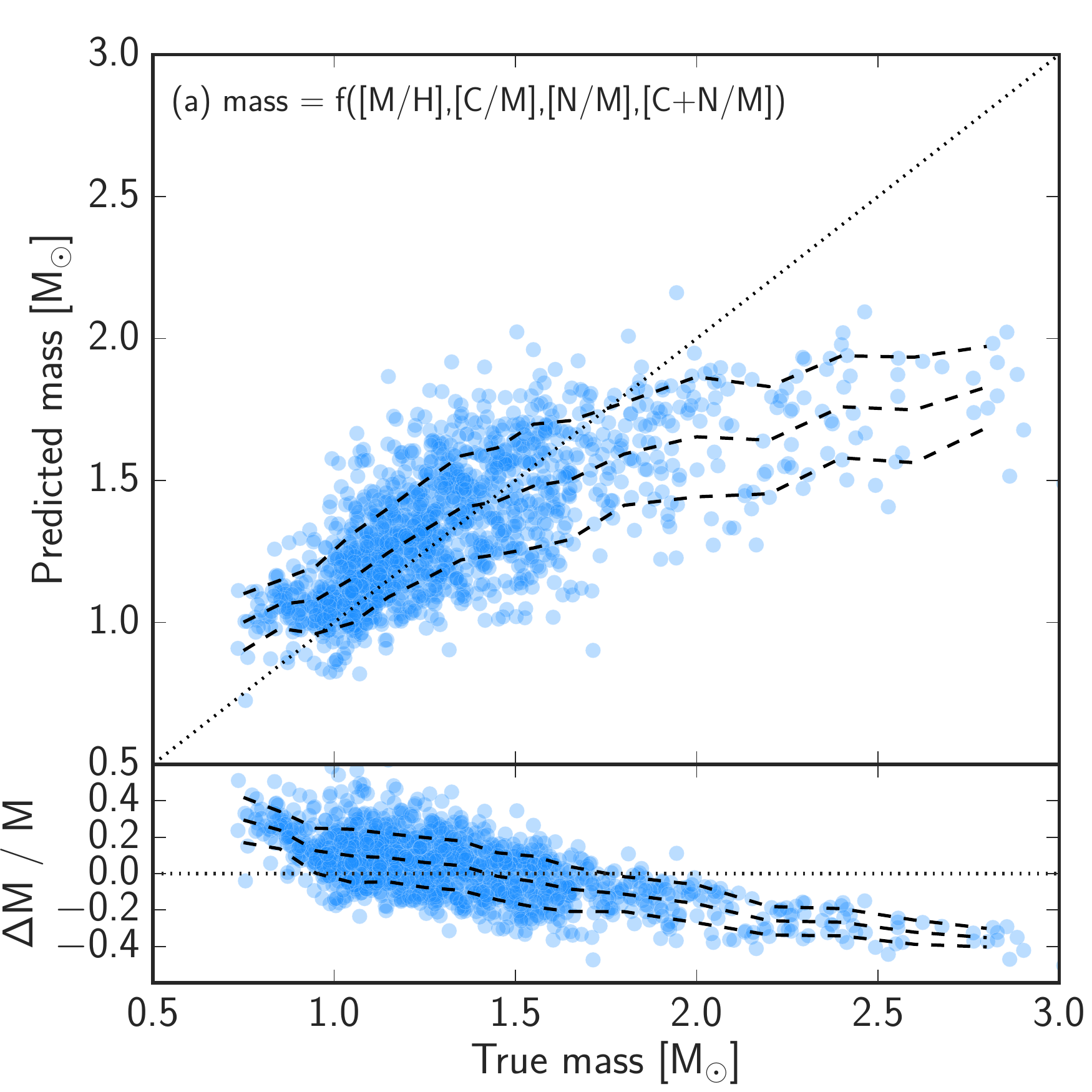}
\includegraphics[width=0.48\textwidth]{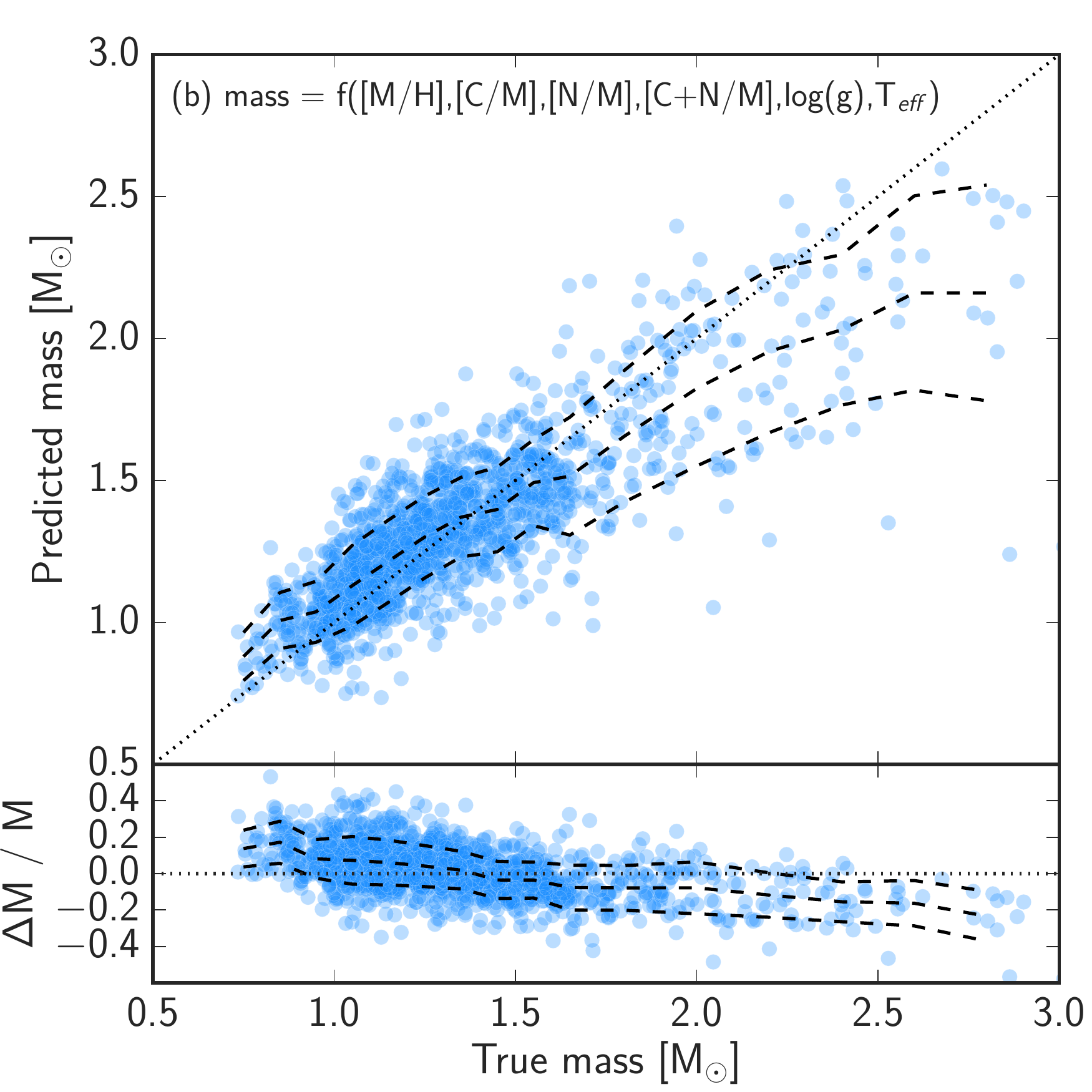}
\caption{Results of the two different mass fits we performed: on the left, the fit including only element abundances, and on the right the fit that also includes \teff and log(g). In both cases, we show on the top the predicted mass as a function of the true mass, the dashed lines represent the mean of the relation and the 1-$\sigma$ range around the mean. The bottom panels contain the relative mass error, with also the mean and 1-$\sigma$ range in dashed lines. While both fits show a similar scatter, adding \teff and log(g) allows to reduce the bias significantly.}
\label{fig:fit_mass}
\end{figure*}

\begin{figure*}
\centering 
\includegraphics[width=0.48\textwidth]{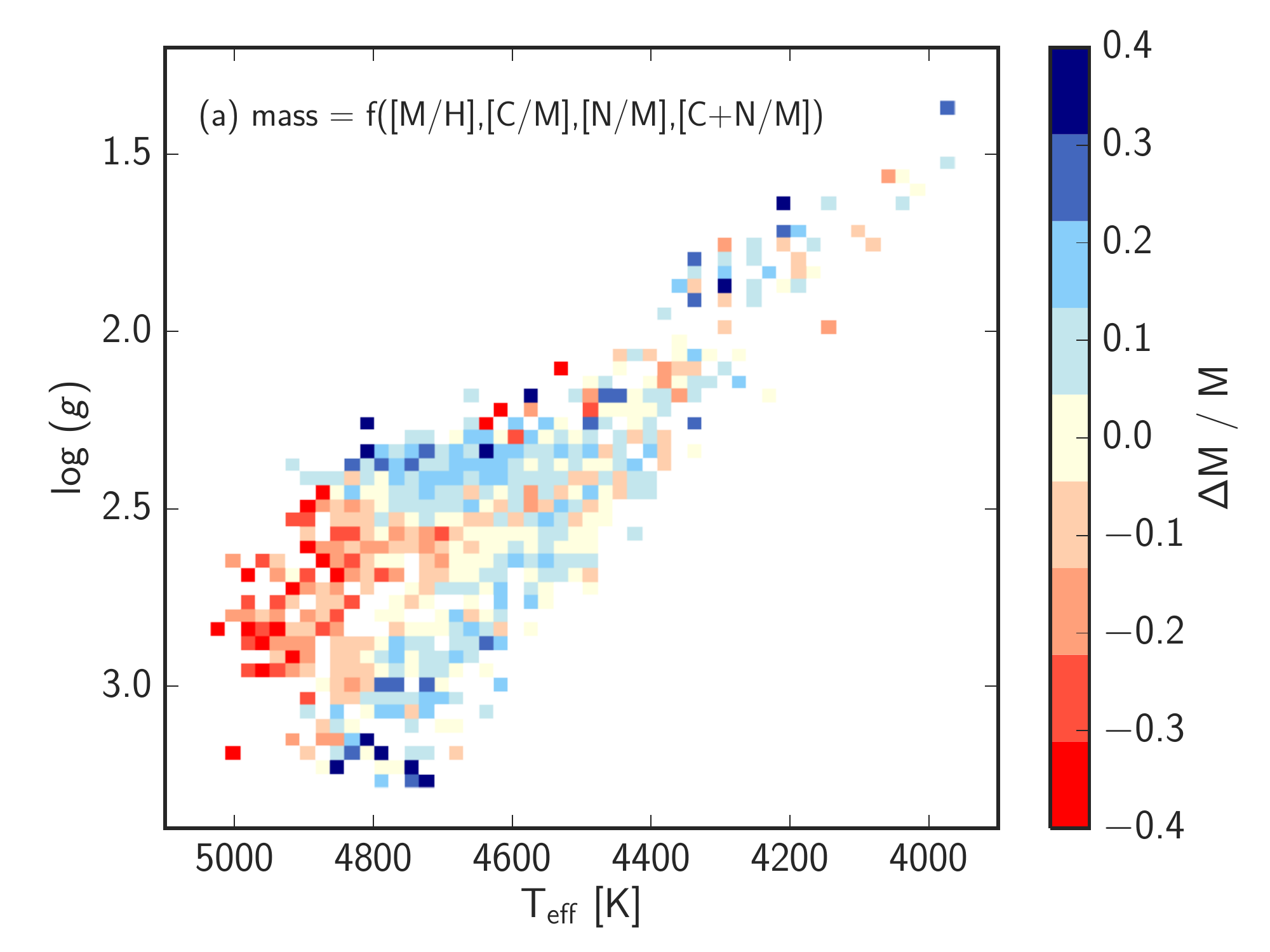}
\includegraphics[width=0.48\textwidth]{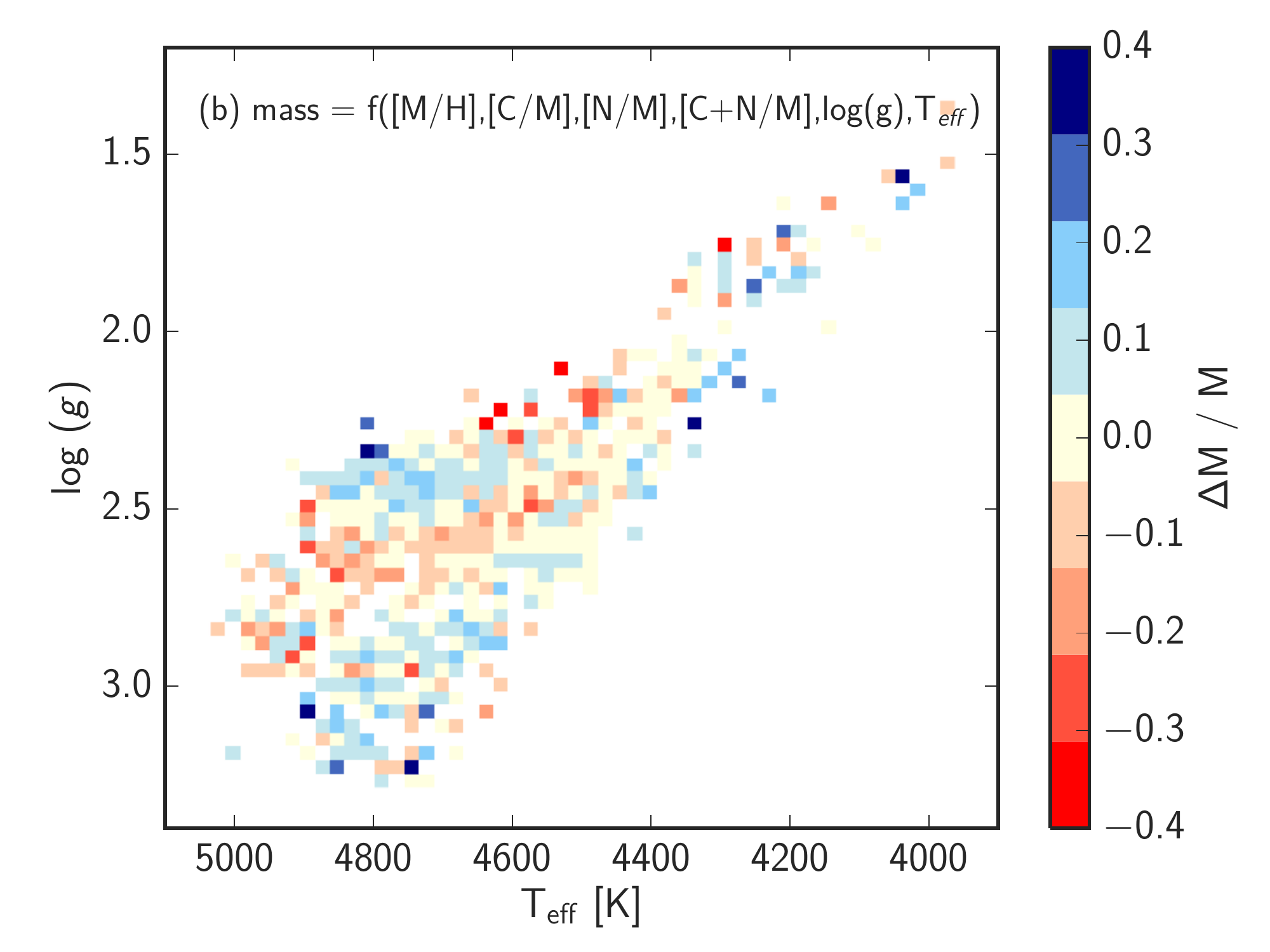}
\caption{Relative mass error for the two different mass fits shown in the log(g) vs \teff plane (the median mass error is shown in bins of log(g) vs \teff -- in red, the model under-predicts the mass, in blue the model over-predicts the mass). The fit that only include element abundances under-predicts the mass of stars in secondary clump (here appearing in red in the left panel at log(g) slightly below 3): these are the high mass stars for which the fit is performing poorly. Adding \teff and log(g) as labels decreases the magnitude of the residuals overall and also makes their distribution more uniform across the H-R diagram.}
\label{fig:fit_residuals}
\end{figure*}

\begin{figure*}
\centering 
\includegraphics[scale=0.35]{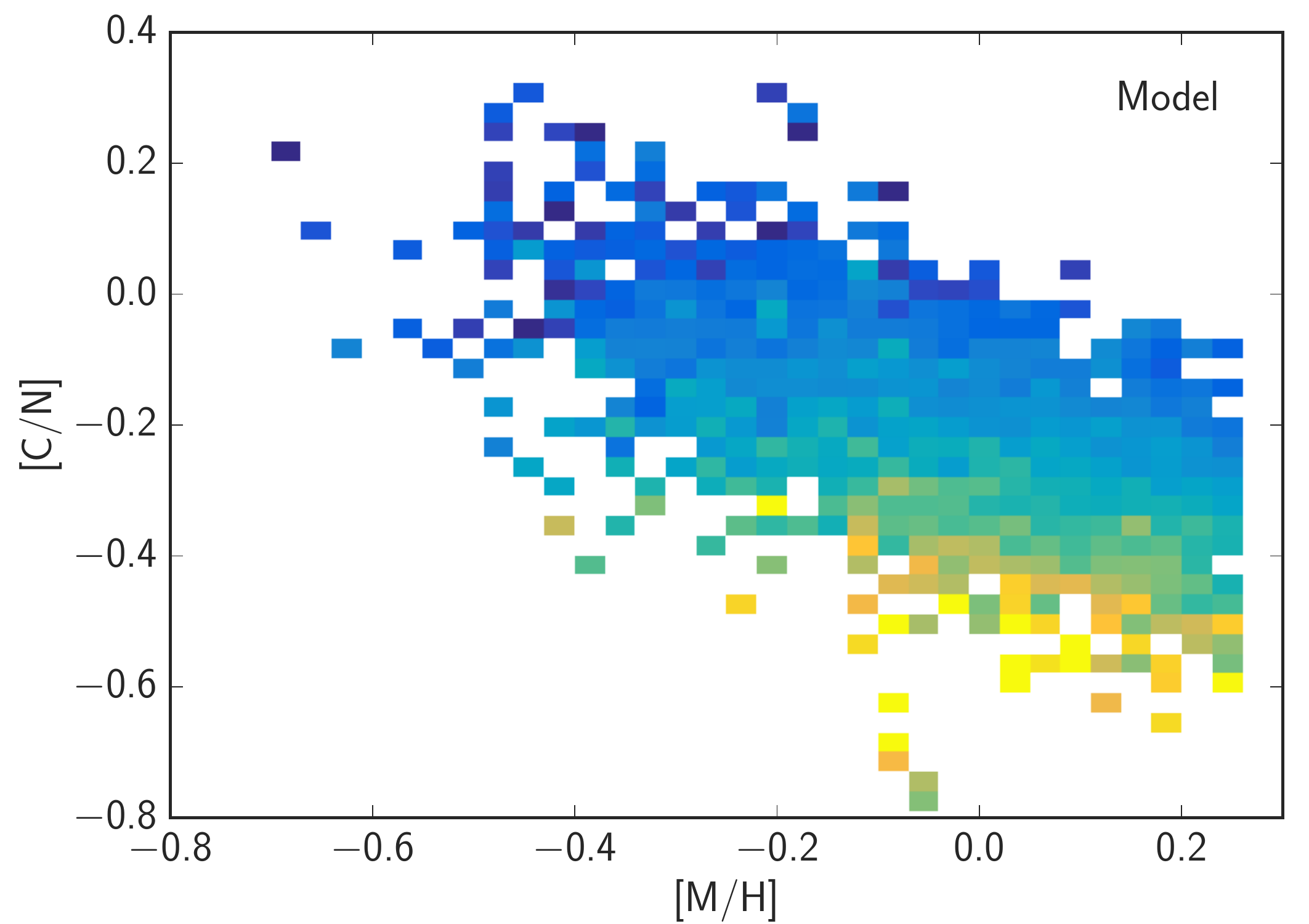}
\includegraphics[scale=0.35]{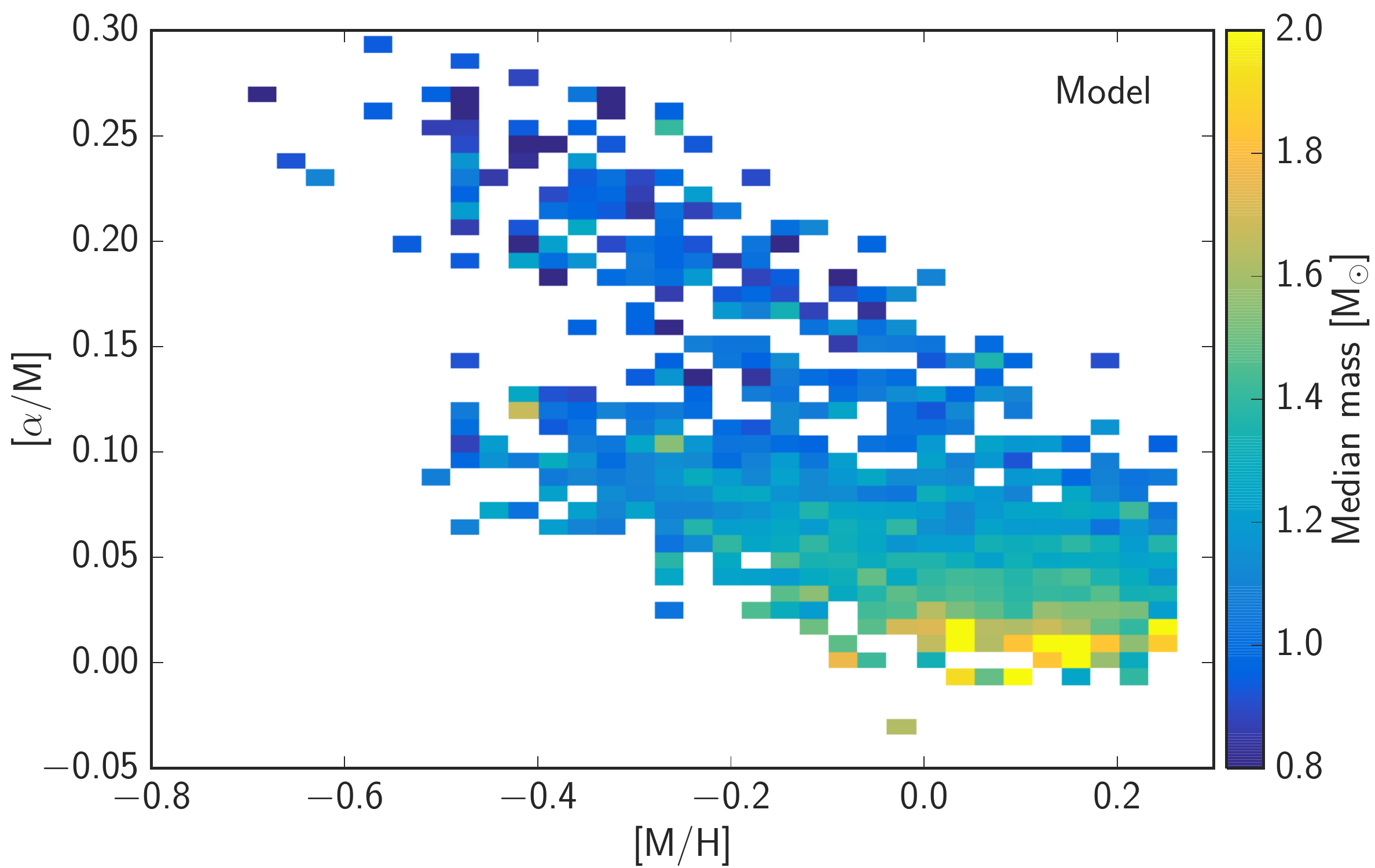}
\caption{Distribution of stars in the [C/N] vs [M/H] plane (left panel) and \aFe vs [M/H] plane (right panel) using the masses predicted from the fit including element abundances, \teff and log(g).}
\label{fig:model_CN_alpha}
\end{figure*}
\subsection{Results}\label{sec:results}

We first apply the method described in the previous section to fit for mass as a function of [M/H], [C/M], [N/M], and [(C+N)/M]. We add [(C+N)/M] in the fit because stellar models predict this remains constant during the dredge-up and is thus characteristic of the birth composition of a star.

The coefficients we obtain for the fitting function are provided in Table \ref{tab:coef_mass1} in Appendix \ref{Appendix_fit}. The mass uncertainty for each star (obtained from 100 different realizations of the model as explained in the previous Section) is of 0.02 \msun on average. This uncertainty is much smaller than the r.m.s error returned by the cross-validation, which is 0.26 \msun, or 18 per cent in fractional error. This means that the individual mass internal uncertainties are meaningless, and that the error budget is dominated by systematic errors (either an inappropriate model, or biases in the data itself).

The left panel of Figure \ref{fig:fit_mass} shows the predicted mass as a function of true mass for the stars in our APOKASC training set, and the relative mass error as a function of the true mass. The dashed lines represent the mean and 1-$\sigma$ scatter around the mean in both cases. The simple fit that we have used performs relatively well for most stars with a mass between 1 and 1.5 \msun but tends to over-predict the mass of low mass stars, and significantly under-predict the mass of massive stars. 

An important aspect to check is if the way we derived the masses themselves could be the source of that bias. There are indeed different ways to determine masses from the seismic parameters: the direct method (used here), and the grid-based method that relies on comparing observed stellar parameters with theoretical isochrones. For the APOKASC sample, both ways of determining masses give similar results, except at very low and very high masses (Figure 3 in \citealp{Martig2015}). This could explain part of the bias we find. However, we have tried to fit for the grid-based masses, and find the same bias to be present and extremely similar.

To explore further the origin of the bias, the left panel of Figure \ref{fig:fit_residuals} shows the individual relative mass error as a function of \teff and log $g$. While the values of the relative errors are small on average, they show some significant structure in the H--R diagram. In particular, the mass of secondary clump stars is systematically under-predicted: these are the stars with a mass of $\sim 2$ \msun, that also appeared as problematic stars in Figure \ref{fig:fit_mass}. 

These massive stars might be outliers in our fits because they actually may not follow the same relation between [C/N] and mass as the rest of the sample. As described in Section \ref{sec:CNOcycles}, massive stars only experience a short RGB phase and do not undergo extra mixing after the first dredge-up. They also do not go through the helium flash at the tip of the RGB \citep{Salaris2005}, they do not shed their envelope away and do not lose as much mass as lower mass stars during this instability phase. These reasons could explain why the mapping of [C/N] to mass could differ for massive stars.

To improve our model, one possibility would be to gather a larger training set, on which we could use more flexible fitting procedures. This is left for future work, as an extended version of the APOKASC sample will be released soon. In the meantime, we try to improve our fit by adding more stellar labels measured by the APOGEE pipeline.

\subsection{Improved fit using \teff and log $g$}\label{sec:improvedFit}

\begin{table*}
\begin{center}
\caption{Stellar parameters for stars in the training set, together with their ``true'' masses and ages as well as the masses and ages predicted by our models. The full table is available in electronic form.}\label{tab:data}
\begin{tabular}{cccccccccc}
\hline
\hline
2MASS ID & \teff[K] & log $g$ & [M/H] & [C/M] &[N/M]& M$_{\mathrm{in}}$[M$_{\odot}$] & M$_{\mathrm{out}}$[M$_{\odot}$]&  age$_{\mathrm{in}}$[Gyr] & age$_{\mathrm{out}}$[Gyr]\\
\hline
2M18583782+4822494 & 4752  & 2.8 &  -0.05 &  -0.15 &  0.20 &  1.49 $\pm$ 0.20  &  1.54 & 2.9 $^{+1.3}_{-0.8}$   &  2.7 \\
2M18571019+4848067 & 4658  & 2.7 &  0.07 &  -0.09 &  0.11 &  1.12 $\pm$ 0.14  &  1.08 & 6.8 $^{+3.0}_{-1.9}$   &  6.9 \\
2M18584464+4857075 & 4499  & 2.7 &  -0.00 &  -0.05 &  0.20 &  1.45 $\pm$ 0.17  &  1.28 & 3.0 $^{+1.3}_{-0.9}$   &  4.5 \\
2M18582108+4901359 & 4169  & 2.1 &  0.04 &  -0.01 &  0.15 &  1.18 $\pm$ 0.16  &  1.29 & 5.7 $^{+4.3}_{-2.1}$   &  5.0 \\
2M18583500+4906208 & 4812  & 3.2 &  0.01 &  -0.07 &  0.21 &  1.45 $\pm$ 0.15  &  1.44 & 3.0 $^{+1.0}_{-0.7}$   &  3.3 \\
2M18581445+4901055 & 4694  & 2.8 &  0.07 &  -0.02 &  0.17 &  1.18 $\pm$ 0.13  &  1.26 & 5.8 $^{+2.4}_{-1.5}$   &  5.1 \\
2M19010271+4837597 & 4555  & 2.5 &  0.24 &  0.00 &  0.16 &  1.13 $\pm$ 0.17  &  0.74 & 6.8 $^{+4.2}_{-2.3}$   &  13.0 \\
2M19005306+4856134 & 4561  & 2.9 &  0.20 &  -0.01 &  0.28 &  1.28 $\pm$ 0.15  &  1.26 & 4.7 $^{+2.9}_{-1.3}$   &  5.2 \\
2M19013400+4908307 & 4748  & 2.9 &  0.04 &  -0.07 &  0.19 &  1.37 $\pm$ 0.15  &  1.47 & 3.6 $^{+1.4}_{-0.9}$   &  3.3 \\
2M19003958+4858122 & 4580  & 2.8 &  0.22 &  -0.05 &  0.27 &  1.34 $\pm$ 0.26  &  1.33 & 4.2 $^{+3.7}_{-1.6}$   &  4.6 \\
\dots \\
\hline
\end{tabular}
\end{center}
\end{table*}
Because the mass residuals show some structure in the HR diagram, we try to include \teff and log $g$  in the fit. This new model is less physically motivated in the sense that \teff and log $g$  do not directly govern the stellar evolution physics explaining why mass is related on [C/M] and [N/M]. 
It could however empirically capture variations in the correlation between mass and abundances as a function of stellar evolutionary phase.
We find that adding these two additional labels leads to better fits to the data: the r.m.s error returned by the cross-validation decreases to 0.21 \msun, or 14 per cent in fractional error (this is again much larger than individual internal mass uncertainties). As for the previous Section, the fit coefficients are given in in Table \ref{tab:coef_mass2} in Appendix \ref{Appendix_fit}.

The right panel of Figure \ref{fig:fit_mass} shows the relation between predicted and true mass for the new model, where we recall that ``true'' mass refers to the seismic estimates. The bias that was present in the previous fit (see right panel of Fig.\,\ref{fig:fit_mass}) is still there, but is strongly reduced, particularly at high masses. The reduction of the bias at high mass is mostly due to the inclusion of \teff while both \teff and log $g$ help for the low mass range. The comparison between the left and the right panel of Figure \ref{fig:fit_residuals}  illustrates that the overall magnitude of the residuals decreases, especially for the secondary clump stars. There are still massive stars for which the model underestimates their masses by 20--30\%: these stars are mostly outside the secondary clump, but at lower log $g$ (these are the yellow dots at log $g<2.5$ in Figure \ref{fig:HRmass}). Some of these stars might have accreted mass from a companion, an event that would have have altered both their mass and their surface abundances.

By contrast, our model tends to over-predict the mass of low mass stars ($\leq 0.9$\,\msun). If we consider stars with masses lower than 0.9 \msun, most of them are located in the red clump in a  very tight range of log $g$ = 2.3--2.4. This range of log $g$ is consistent with theoretical expectations for old stars as shown in Figure 11 of \cite{Martig2015}. For some of these stars, the mass is correctly predicted by our models, while other stars have a mass error of 30--40\%. We suspect that such a scatter could be due to different mass loss rates undergone by the stars during the RGB phase. Low-mass stars are indeed the ones for which mass loss is the strongest (see Figure \ref{fig:dredgeup}). As a result of loosing a significant amount of mass during the RGB phase, their [C/N] ratio would be  consistent with a higher mass than their actual present day mass. The scatter in mass loss rates could partly be due to tidally enhanced stellar winds in stars with a binary companion \citep{Tout1988,Lei2013}. 

Overall, these biases in the low and high mass ranges result in a larger r.m.s. mass error for stars in the RC (0.24 \msun) compared to RGB stars (0.15 \msun).

\begin{figure}
\centering 
\includegraphics[width=0.48\textwidth]{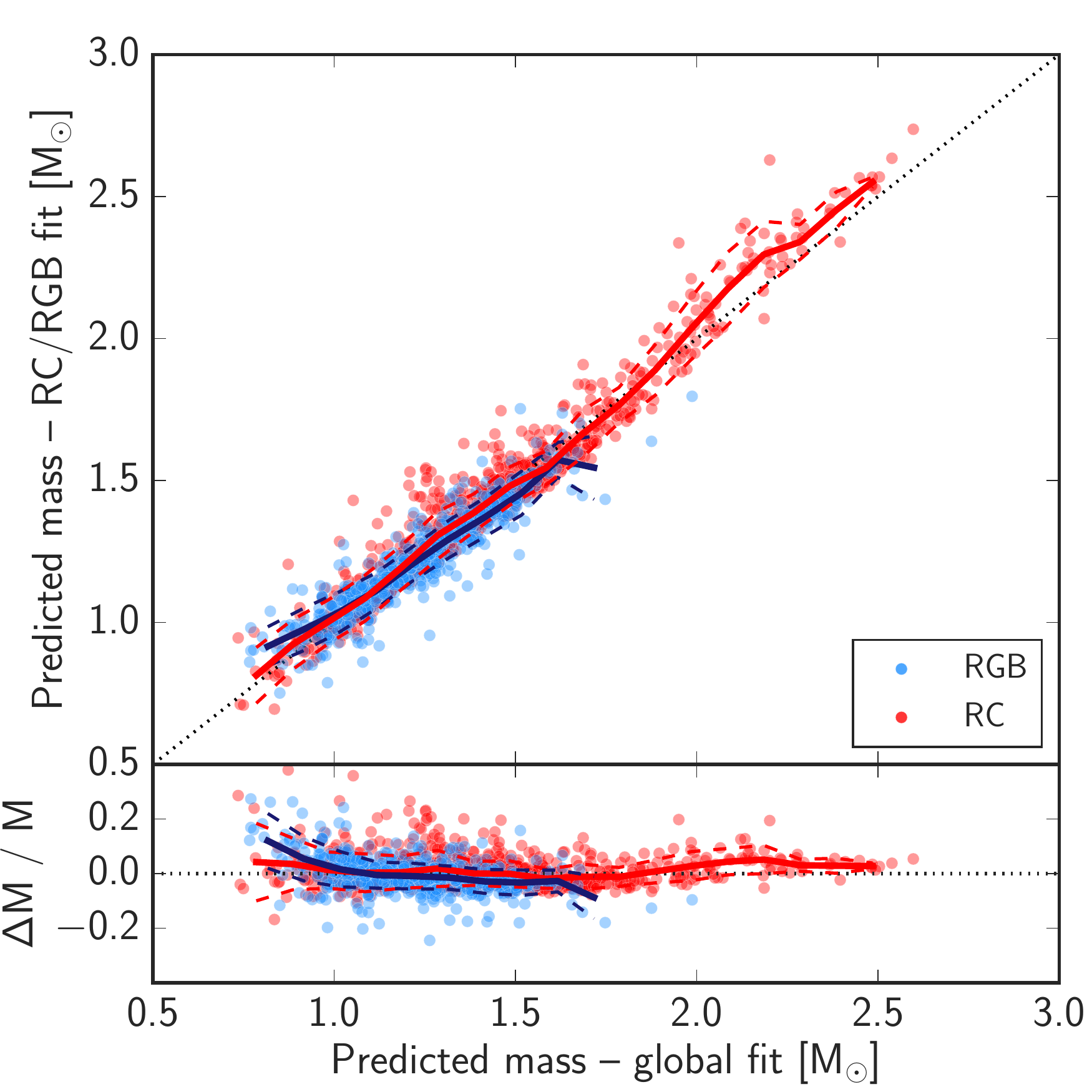}
\caption{Comparison of masses obtained when the RC and RGB sample are fitted separately to masses obtained from the global fit to all stars together. This shows that both methods give very similar results, with a scatter of 5\% for both RGB and RC stars.}
\label{fig:rc_rgb}
\end{figure}

To test whether the biases could be due to a different scaling between mass and abundances for RGB and RC stars (that maybe would not be captured by the inclusion of \teff and log $g$ in the fit), we compare the masses we predict with the global fit to masses that are obtained from a separate fit to the RGB and RC stars. Figure \ref{fig:rc_rgb} shows that the predicted masses are very similar in both cases, so that the systematic biases we find are not eliminated by fitting RC and RGB separately.

In spite of these residual biases at small and high mass, the fit is successful at reproducing most of the global trends seen in the data
Figure \ref{fig:model_CN_alpha} shows the distribution of the fitted masses in the [C/N] vs [M/H] and \aFe vs [M/H] planes, highlighting the consistency of the model with the data.

\subsection{Fitting for age}\label{sec:agefitting}

\begin{figure*}
\centering 
\includegraphics[width=0.48\textwidth]{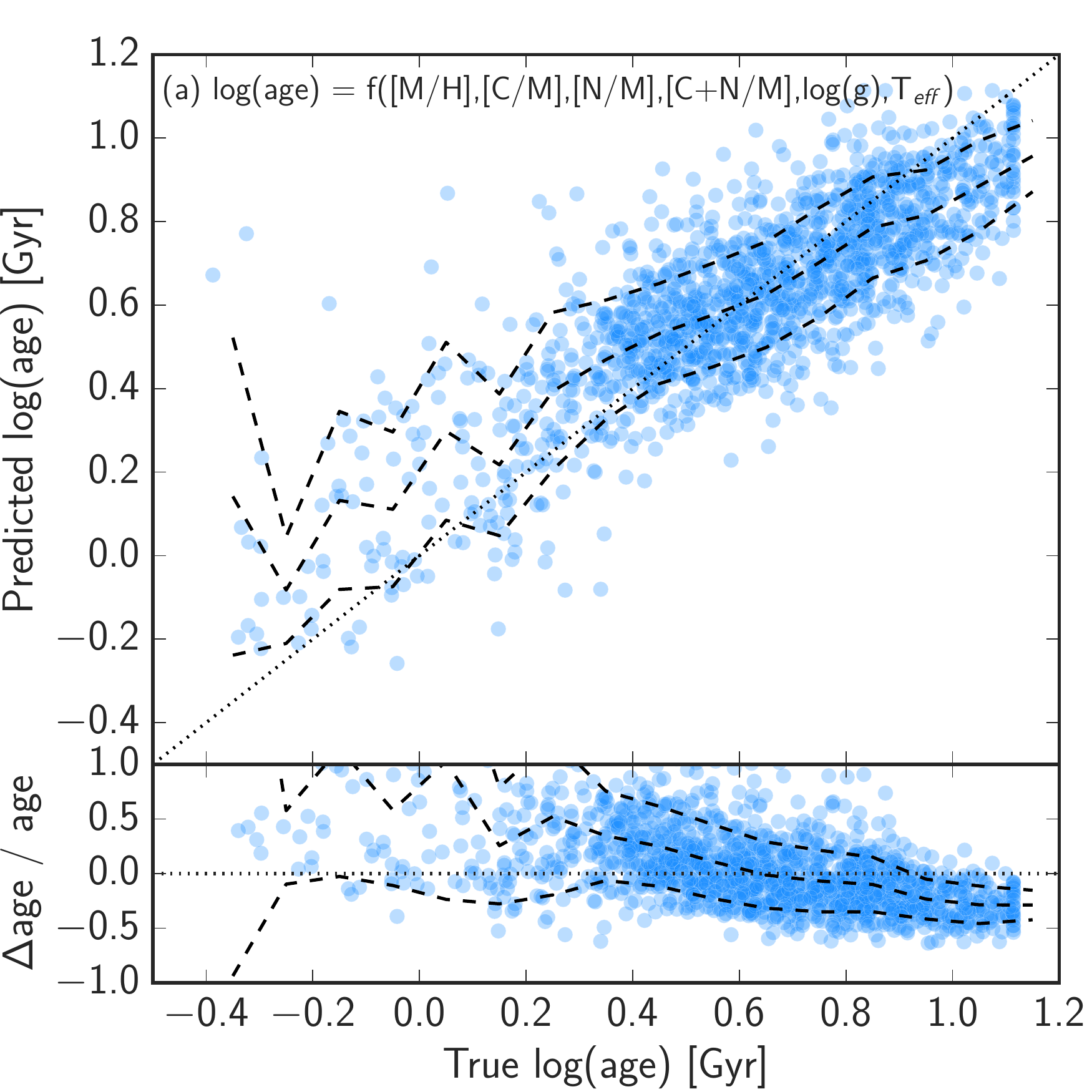}
\includegraphics[width=0.48\textwidth]{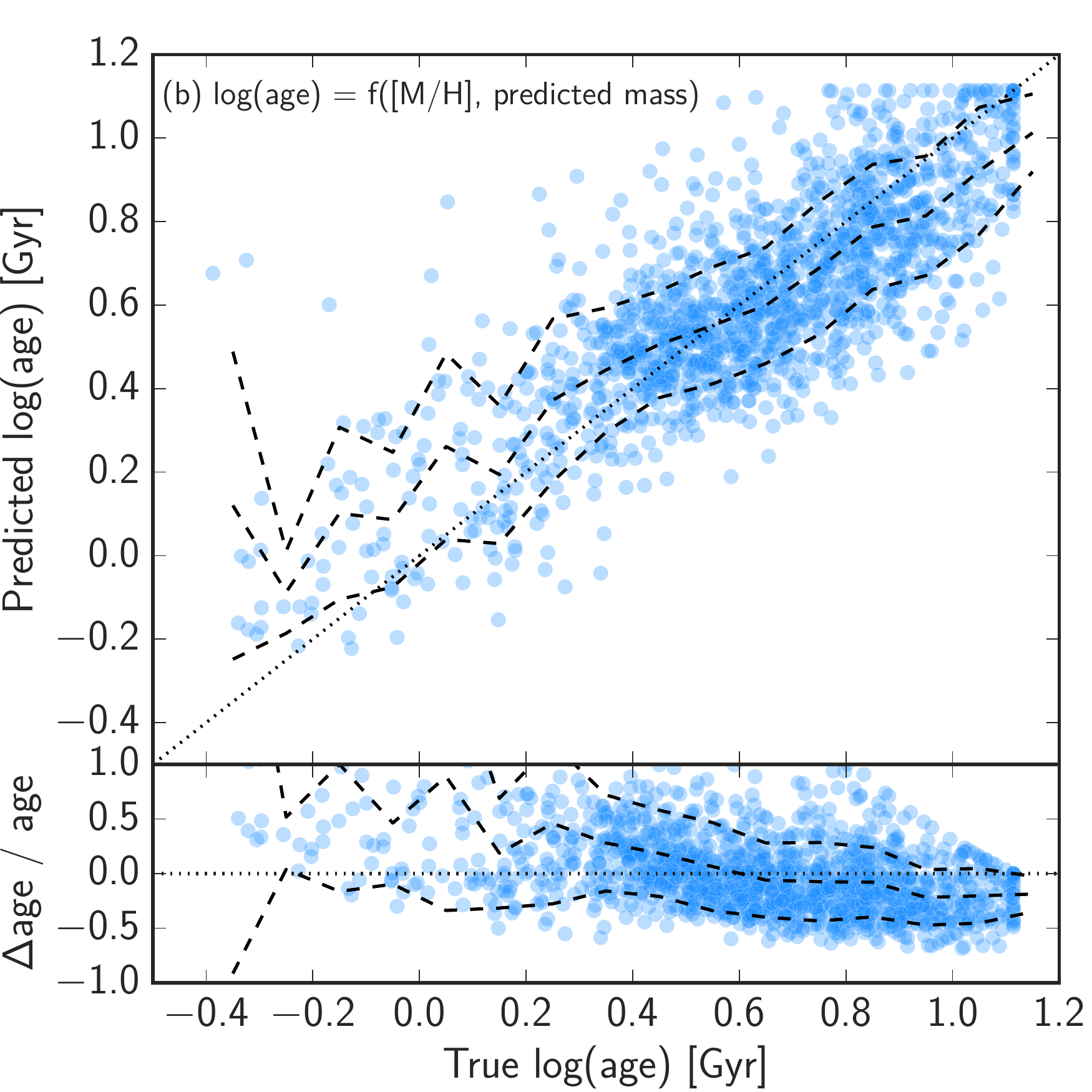}
\caption{Results of the two different ways of determining ages: on the left, log(age) is directly obtained from a fit to element abundances, \teff and log $g$, while on the right log(age) is derived from [M/H] and the predicted mass, as described in Section 3.6. In both cases, we show on the top the predicted log(age) as a function of the true log(age), the dashed lines represent the mean of the relation and the 1-$\sigma$ range around the mean. The bottom panels contain the relative age error, with also the mean and 1-$\sigma$ range in dashed lines.}
\label{fig:fit_age}
\end{figure*}
Given the success of our model to predict masses, we also apply the same technique  to obtain a model that predicts ages using the same set of labels (incl. \teff and log(g)). While age is not the fundamental stellar property that governs the changes in surface abundances on the giant branch, the tight relation between mass and age makes it possible to derive ages from our set of labels. We actually fit for log(age) instead of age, to ensure that the fitted ages are positive. We also impose an upper limit of 13 Gyr to the ages we derive. The coefficients and their errors are given in Table \ref{tab:coef_age} in Appendix \ref{Appendix_fit}.

The cross validation algorithm gives an absolute r.m.s age error of 1.9 Gyr, and 40 per cent relative error  (the relative age error is only computed for stars older than 1.5 Gyr since younger stars have a much greater relative age error).

In Figure \ref{fig:fit_age}, we show the result of the fit on the left panel, and on the right panel the ages we would obtain by translating the fitted masses into ages using the procedure described in Section \ref{sec:sample}. Both ways of determining ages give similar results in term of general bias, with a slightly smaller scatter if ages are fitted directly. The biases we find here in the age fit are directly linked to the fits we had in the mass fits: the ages are under-predicted at high age and over-predicted at low age.

However, and in spite of their relatively small associated errors, these ages have to be taken carefully. They do have a very substantial  model-dependence, especially for red clump stars, where the relation between mass and age strongly depends on the mass-loss prescription. We encourage the readers to use our predicted masses and to convert them into age themselves based on their own favourite stellar evolution model.

\section{Application to DR12 data: deriving mass and age for large samples of stars}
\label{sec:dr12}

In this Section, we apply our model to APOGEE DR12 data for which no prior information is available from Kepler, in particular age information. 
Our model allows us to transfer information from the sample with asteroseismic data to a much larger sample. However, in this present paper, we only aim to demonstrate astrophysical plausibility of our results and we leave a detailed discussion of the age structure of the Milky Way to future papers.

\subsection{A word of caution: stellar evolution vs. galactic chemical evolution}

\begin{figure}
\centering 
\includegraphics[width=0.5\textwidth]{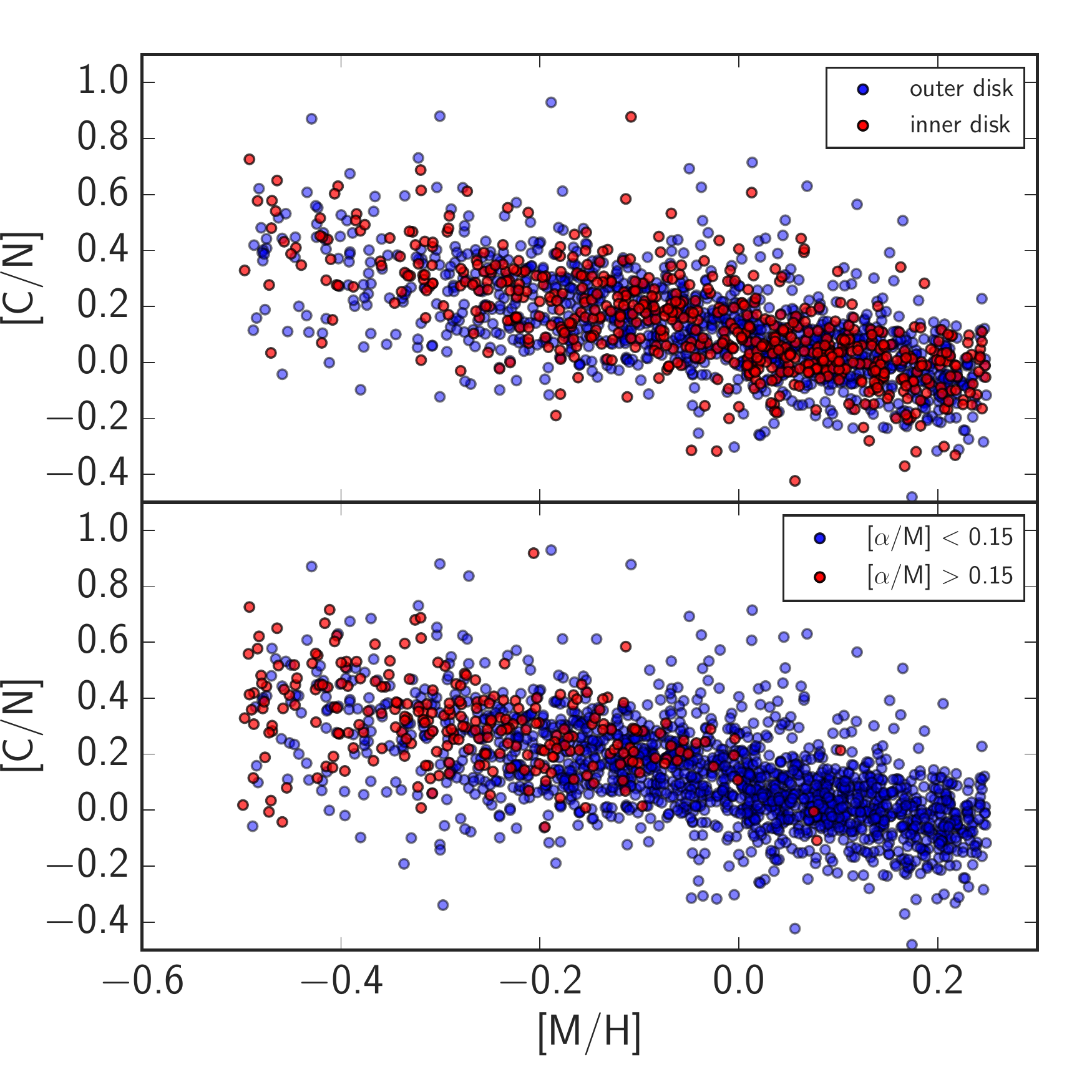}
\caption{[C/N] ratio as a function of metallicity for a sample of 1,943 pre dredge-up giants in DR12 (selected as described in the text and in Figure \ref{fig:selection}). Since these stars have not been through the dredge-up yet, their surface abundances reflect their birth properties. The top panel compares stars in the inner and outer disks (i.e., galactocentric distance smaller or greater than 8 kpc, with an additional cut to only keep stars within 1~kpc of the mid-plane), the bottom panel compares $\alpha$-rich and $\alpha$-poor stars. This shows that galactic chemical evolution does not affect the shape of the [C/N] vs. [M/H] relation in the range of distances probed by this sample. The relation is also the same for  $\alpha$-rich and $\alpha$-poor stars.}
\label{fig:subgiants}
\end{figure}
Because of potential disagreements between measurements of carbon and nitrogen abundances between different surveys, the fitting functions we provide are only applicable to APOGEE DR12 data. The method remains valid, but the models would need to be re-calibrated for any different survey, or even for future data releases of APOGEE. 

Even when only applying the fits to APOGEE DR12, a complication comes from the fact that the stars' C and N abundances might reflect both stellar evolution and initial abundances in the stars at birth. For samples covering large portions of the Milky Way, one could imagine that the variations of birth abundances from one region to another could become significant. Such variations might create fake spatial trends in the derived masses and ages. 

To study the birth abundances of stars, one needs a sample of stars on the main sequence or on the subgiant branch, i.e., stars that have not gone through the first dredge-up yet. Stellar parameters for dwarfs have to be taken with extreme caution in DR12 because the spectral grids used to fit the observed spectra do not include rotation \citep[see][]{Holtzman2015}. We identified instead a sample of 1,943 giants or subgiants with surface abundances of C and N that are consistent with a pre dredge-up composition. These stars are identified as being at the very bottom of the RGB, and as having a high [C/N] ratio (see the black box in Figure \ref{fig:selection}). The cuts we use to define the pre dredge-up sample  are the following:

\begin{equation}
  \begin{cases}
  T_{\mathrm{eff}} < 5200\\
  3.5 < \log(g) < 4\\
0.00125 \times T_{\mathrm{eff}} -2.875 <  \log(g) < 0.002\times T_{\mathrm{eff}} - 6.0
  \end{cases}
\end{equation}

The [C/N] ratio for these stars reflects how chemical evolution proceeds in the Milky Way. We show in Figure \ref{fig:subgiants} the relation between [C/N] and [M/H] for the subgiants as a function of their spatial location (top panel, using distances from Ness et al., in prep) and content in $\alpha$ elements (bottom panel). This shows that for this sample of stars, the relation between birth [C/N] and metallicity is independent of location within the Milky Way (within the range of distances probed by the subgiants, which is unfortunately limited to a few kpc around the sun).

The study of the carbon and nitrogen abundances of pre dredge-up giants shows that galactic chemical evolution proceeds in the same way over the range of distances these stars probe, so that our fits are there directly applicable to measure ages and masses. Special caution should be taken when applying the fits in regions of the Milky Way where chemical evolution could be more complex, like in the bulge/bar region.
We limit the possibility of such an effect by limiting the fits to stars within the same range of [(C+N)/M] as our APOKASC training set. By selecting DR12 stars in the same range of [(C+N)/M] as the APOKASC sample, we automatically select stars within the same range of birth abundances. We also include [(C+N)/M] as an input label in the fits to capture any dependence of the predicted mass on this parameter.

\subsection{Stellar masses and ages for APOGEE DR12 stars}

\begin{table*}
\begin{center}
\caption{Predicted masses and ages for stars in APOGEE DR12. We do not provide individual mass and age uncertainties because the error budget is dominated by systematic errors. The full table is available in electronic form.}\label{tab:data_dr12}
\begin{tabular}{cccccccc}
\hline
\hline
2MASS ID & \teff[K] & log $g$ & [M/H] & [C/M] &[N/M]& M$_{\mathrm{out}}$ [M$_{\odot}$] & age$_{\mathrm{out}}$ [Gyr]\\
\hline
2M00000211+6327470 & 4600  & 2.5 &  0.02 &  -0.20 &  0.28 &   1.53 & 2.9\\
2M00000446+5854329 & 4725  & 2.9 &  0.02 &  -0.05 &  0.19 &   1.41 & 3.6\\
2M00000535+1504343 & 4791  & 3.3 &  -0.06 &  0.01 &  0.06 &   1.09 & 7.5\\
2M00000797+6436119 & 4449  & 2.5 &  -0.21 &  -0.05 &  0.18 &   1.29 & 3.9\\
2M00000818+5634264 & 4895  & 2.9 &  -0.19 &  0.10 &  -0.02 &   1.31 & 4.9\\
2M00000866+7122144 & 4585  & 2.7 &  -0.07 &  -0.09 &  0.25 &   1.40 & 3.2\\
2M00001104+6348085 & 4865  & 3.3 &  0.06 &  -0.09 &  0.15 &   1.57 & 2.8\\
2M00001242+5524391 & 4579  & 2.6 &  0.12 &  -0.01 &  0.25 &   1.17 & 6.3\\
2M00001296+5851378 & 4659  & 2.9 &  0.07 &  0.06 &  0.19 &   1.25 & 5.1\\
2M00001328+5725563 & 4461  & 2.6 &  0.10 &  -0.08 &  0.25 &   1.36 & 3.9\\
\dots \\
\hline
\end{tabular}
\end{center}
\end{table*}

\begin{figure*}
\centering 
\includegraphics[width=0.48\textwidth]{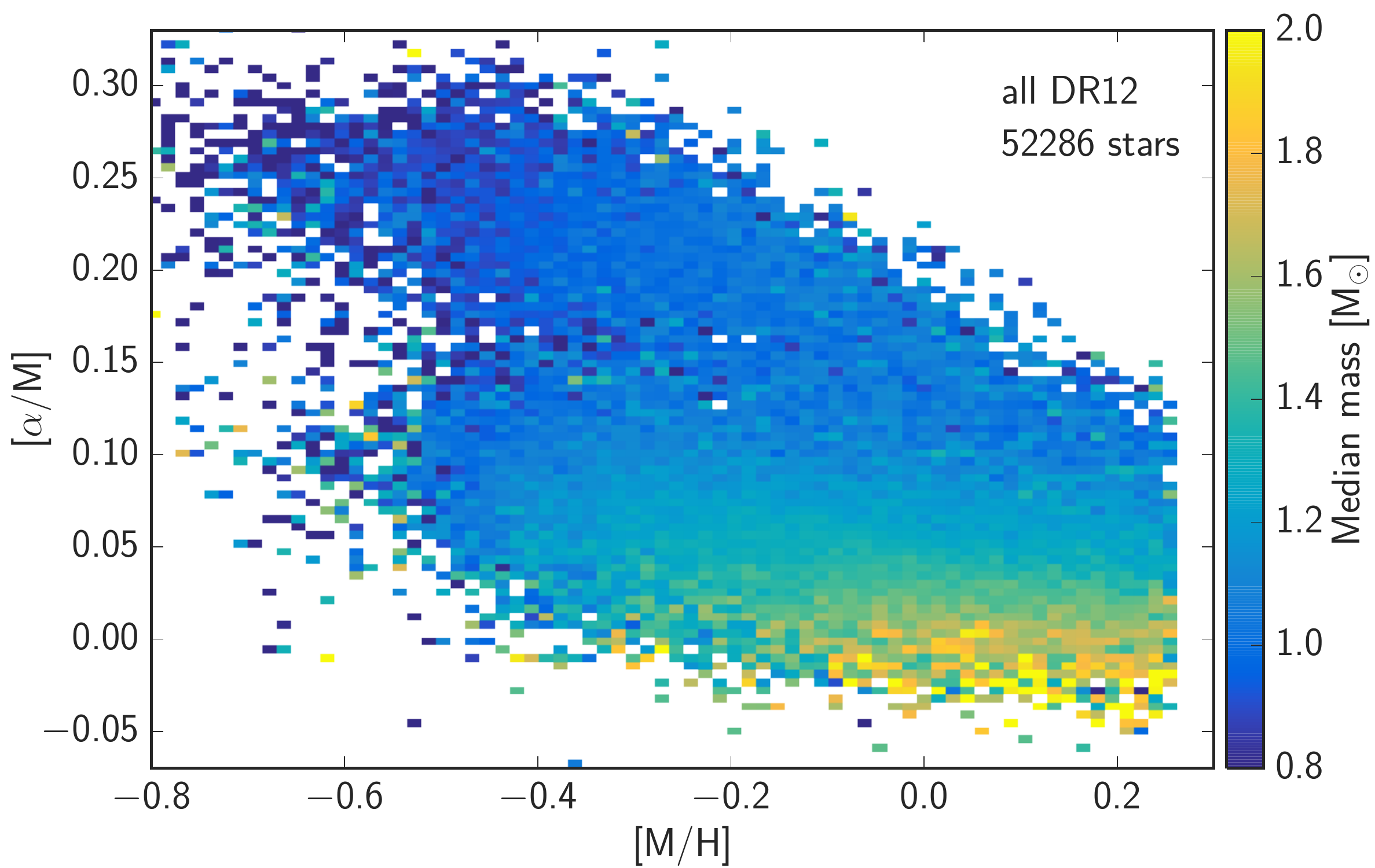}
\includegraphics[width=0.48\textwidth]{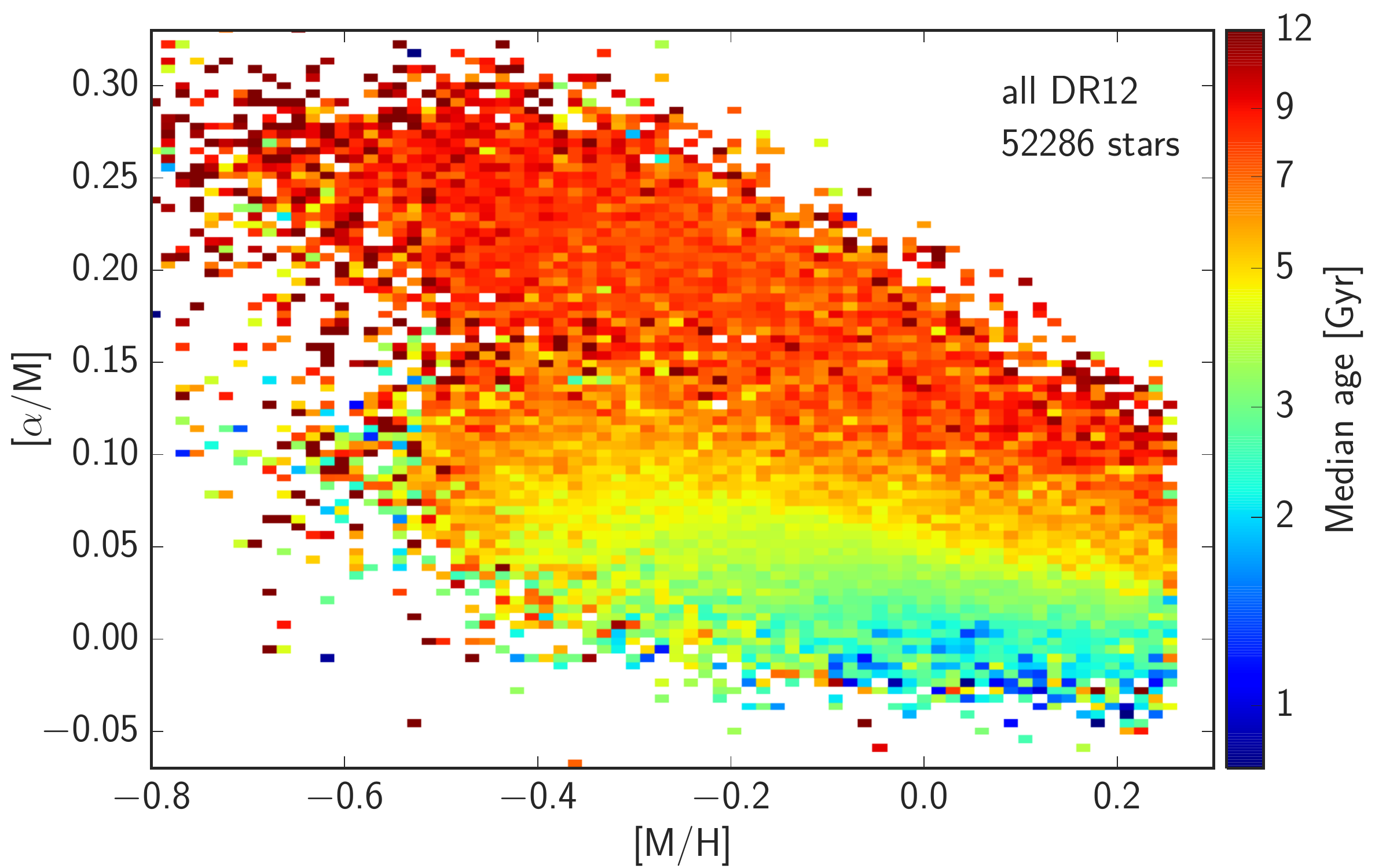}
\caption{Application of our model to APOGEE DR12 data. Both panels represent \aFe as a function of [M/H], colour-coded by predicted mass on the left and predicted age on the right}
\label{fig:dr12}
\end{figure*}

To apply our model to the whole DR12 dataset, we first apply the same quality cuts to DR12 as the ones we described in Section \ref{sec:sample}, and then do some additional cuts to ensure that we are not extrapolating results into regions of the parameter space not covered by our APOKASC sample. These cuts are the following (as a reminder, all parameter values mentioned here are the ones found in the FPARAM array):

\begin{equation}
  \begin{cases}
  [M/H] > -0.8\\
  4000 < T_{eff} < 5000\\
  1.8 < \log(g) < 3.3\\
  -0.25 <[C/M]  < 0.15\\
  -0.1 < [N/M]< 0.45\\
   -0.1  < [(C+N)/M] < 0.15\\
    -0.6 < [C/N] < 0.2 \\
  \end{cases}
\end{equation}
The cut on log~$g$ is also important to ensure that we only include post dredge-up giants, for which the correlation between mass and [C/N] is in place. 
52,286 stars remain after the cuts; their resulting masses and ages are given in Table \ref{tab:data_dr12} and are shown in Figure \ref{fig:dr12} in the \aFe versus [M/H] plane. 

Even though \aFe is not included in our fits, we naturally recover the trend of \aFe vs age that is expected from studies in the solar neighbourhood \citep{Fuhrmann2011,Haywood2013, Bensby2014, Bergemann2014}. We find that the $\alpha$--rich sequence is significantly older than the $\alpha$--poor sequence. However, our analysis does not support the idea of a clear age discontinuity between thin and thick disk (as was argued by \citealp{Masseron2015} based on the difference of [C/N] for the two components). We will explore this aspect in more detail in future papers.
Finally, some outliers appear on top of the mean relation apparent in Figure \ref{fig:dr12}: some of the $\alpha$--rich stars are young, some of the alpha-poor stars are old. If real, these stars would provide interesting constraints to models of radial mixing and Galactic chemical evolution \citep[][]{Martig2015, Chiappini2015}. However, these stars could also be catastrophic outliers in our fits, and their ages would need to be independently confirmed with other techniques before we can draw any conclusions about them.

It is also important to mention that while the relative distribution of ages looks plausible, the absolute scaling might be slightly off. Stars with \aFe$>0.15$ here have a median age of 7.9 Gyr, while previous studies suggest ages of the order of 9--10 Gyr for the $\alpha$-rich sequence \citep{Haywood2013, Bensby2014, Bergemann2014}. Part of an explanation for the too low median age of the $\alpha$-rich stars could be related to the cuts we have to apply to the DR12 sample (that might remove part of the parameter space where older stars would be found), but this issue might simply reflect the fact that our model is known to underestimate the ages of old stars (as shown in Figure \ref{fig:fit_age}).

An important sub-sample of DR12 is the red clump catalogue of \cite{Bovy2014}. A first advantage is that selecting stars of a given evolutionary stage should reduce biases in our relative mass determination, even though ages for the RC are very dependent on the mass loss prescription adopted. Another important advantage of that RC sample is that distances have been determined with an individual uncertainty of 5\%, which allows to study the spatial distribution of stars as a function of their age.

\begin{figure}
\centering 
\includegraphics[width=0.48\textwidth]{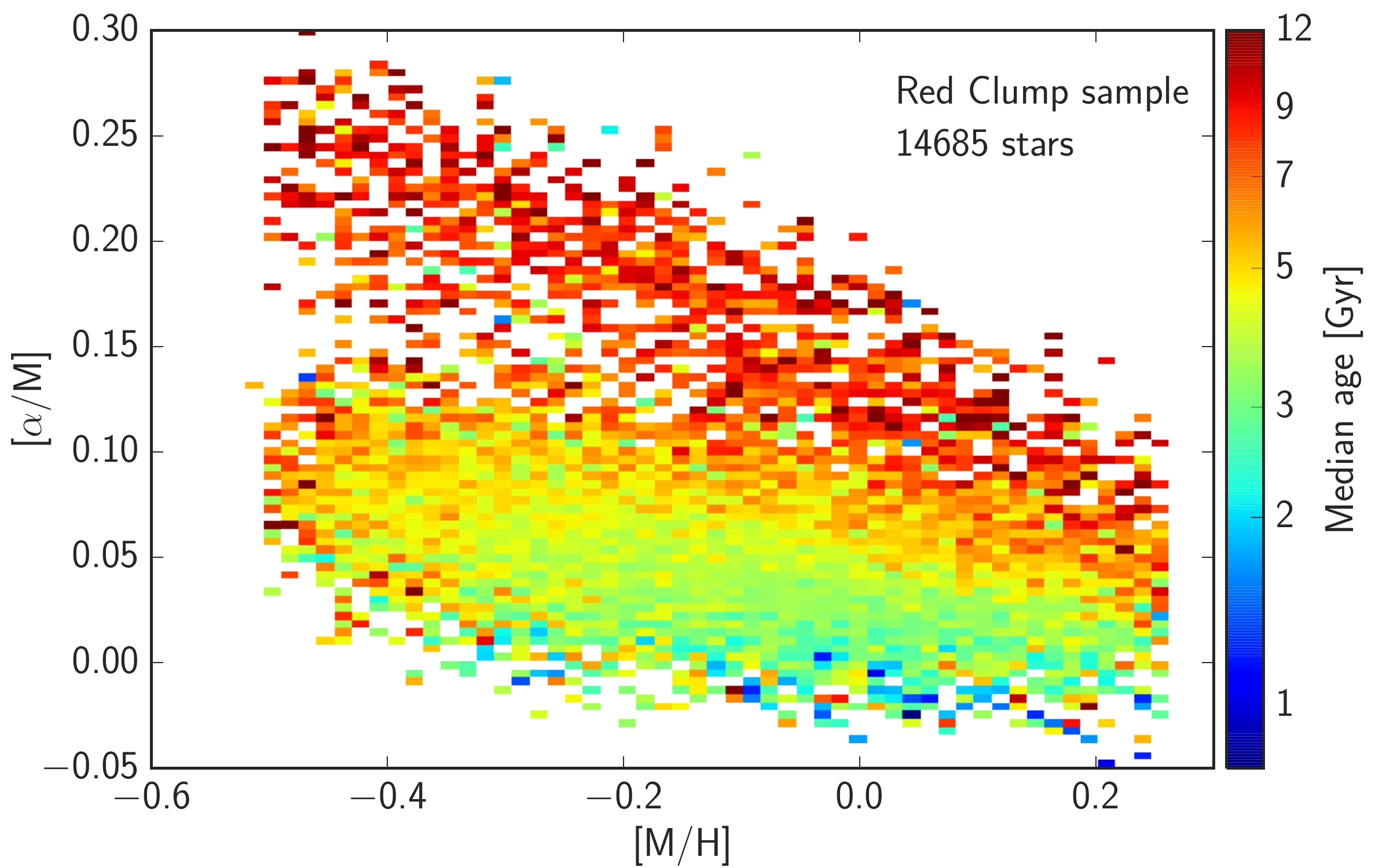}
\caption{Application of our model to the red clump catalogue of Bovy et al. (2014). The distribution of stellar ages in the \aFe vs [M/H] plane is consistent with the larger DR12 sample shown in Figure \ref{fig:dr12}.}
\label{fig:rc_alpha}
\end{figure}

\begin{figure}
\centering 
\includegraphics[width=0.48\textwidth]{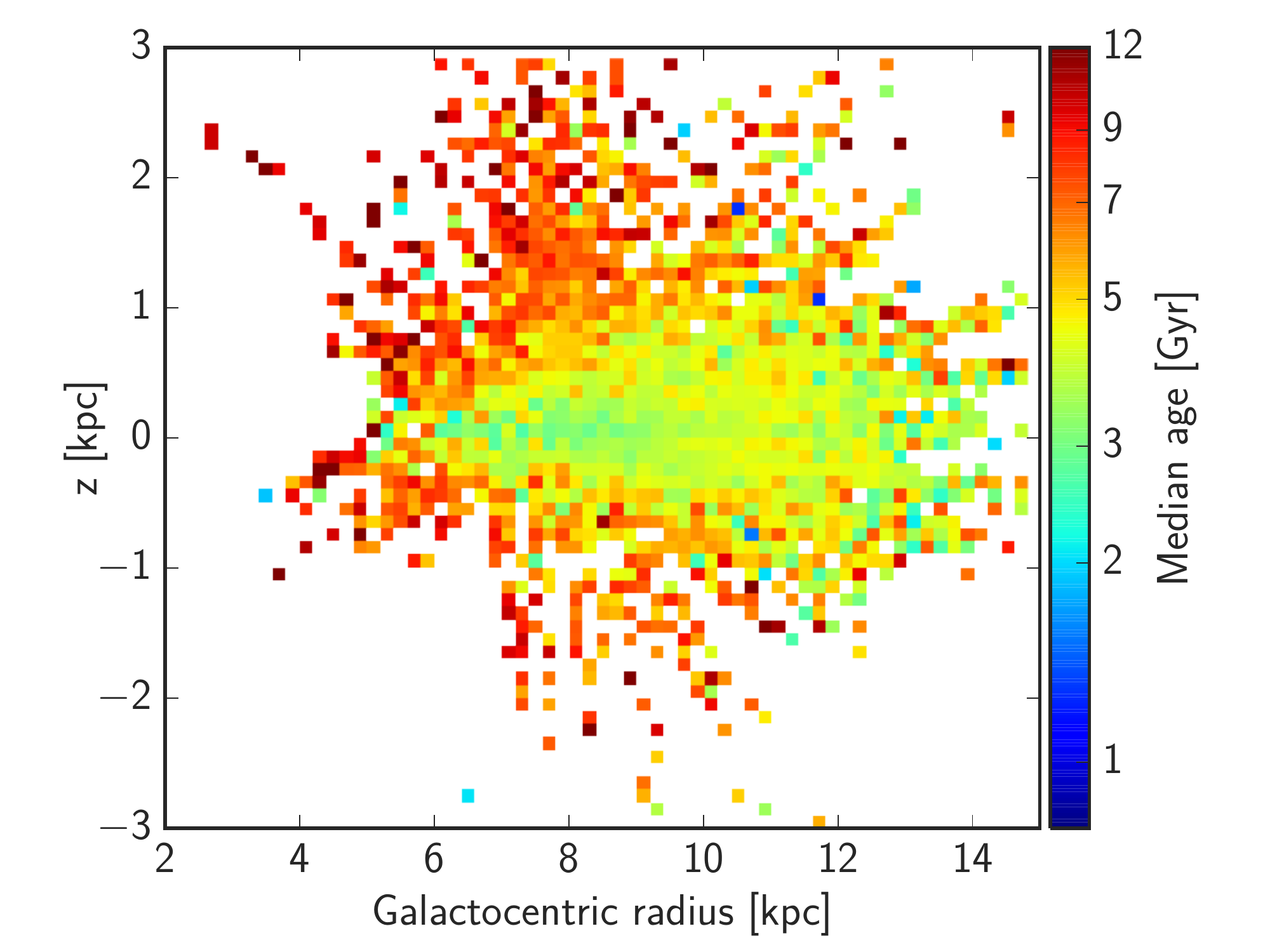}
\caption{Spatial distribution of stars in the red clump catalogue, color coded by median age in bins of galactocentric radius and height above and below the mid-plane. The young stars are found close to the mid-plane, while old stars extend much further above and below the disk.}
\label{fig:rc_rz}
\end{figure}

Applying our cuts to the RC catalogue produces a sample of 14,685 stars.
Figure \ref{fig:rc_alpha} represents the age distribution of these stars in the \aFe vs [M/H] plane, showing results consistent with the larger DR12 sample. Figure \ref{fig:rc_rz} shows the spatial distribution of stars colour-coded by their age. As expected, young stars are concentrated towards the disk mid-plane and older stars extend to higher height above and below the disk. The existence of such spatial correlations reinforces the plausibility of our ages, at least in a relative sense.


\section{Conclusion}
\label{sec:conclusion}

We have laid out a powerful and practical approach to estimate stellar masses, and implied ages, for giant stars on the basis of the stellar labels derived from their spectra. We use a sample of 1,475 giant stars with asteroseismic mass estimates from the APOKASC survey to study and model the correlation between stellar mass and surface abundances of carbon and nitrogen. The power of our approach is that for the first time it is possible to empirically link mass and C, N abundances for a large sample of stars, instead of relying on models to make the connection between both \citep[as was done for instance by][]{Masseron2015}. 
We show that, as expected from stellar evolution models, the [C/N] ratio of giants decreases with increasing stellar mass. The magnitude of the observed decrease is to first order consistent with simple dredge-up models: we do not see any strong evidence for extra mixing in the APOKASC giants. To further test models of mixing processes would require a sample of stars reaching lower log $g$ and/or lower metallicity, for which these effects might be stronger  \citep{Gratton2000,Spite2005}.  

Using APOKASC as a training set, we provide several sets of fitted formulae to predict mass and age as a function of [M/H], [C/M], [N/M], [(C+N)/M], \teff and log $g$. For the stars in the training set, our models are able to predict masses with relative r.m.s. errors of 14 per cent and r.m.s age errors of 40 per cent. This simple model has a small bias in its mass estimates: our predicted mass are too high at low masses and too low at high masses. This could either mean that our models are not flexible enough, or that the input data contain biases (either in the APOGEE stellar parameters or in the seismic masses), or that the biases reflect different physical scalings between stellar mass and surface abundances for different types of stars. As discussed in Section 5, mixing processes and mass loss efficiency vary as a function of stellar mass and could create part of the bias we observe. Future versions of the APOKASC sample will contain thousands of more stars, including stars at lower metallicities and lower log $g$. This  opens up many possibilities for new projects, including for instance detailed comparisons between stellar models and data, and fitting the data with more flexible methods, such as Gaussian processes. 

We must emphasize that individual mass estimates (and even more so
age estimates) must be viewed with great caution, especially if they seem exceptional. For individual stars, the surface abundance of C and N might not always reflect their present day stellar mass, for instance if the presence of a binary companion altered their surface composition and/or their mass. Our method is therefore perhaps best suited for statistical studies of large samples of stars, and to compare the properties of different populations.

Generally speaking, our method of deriving masses and ages for giants has many advantages. First, it is calibrated on asteroseismic data, and provides a relatively simple prescription to transfer the seismic information onto larger data sets. The ideal situation would be to directly have seismic masses measured for large sample of stars covering a large fraction of the Milky Way, but this is not presently the case. In addition, we also note that relying on \aFe as a proxy for age (or using mono-abundance populations in the \aFe vs [M/H] plane as approximations of mono-age populations) might work to some degree (as we also showed in Section 6), but \aFe is an age indicator that depends on the chemical evolution of the Milky Way, and not on the properties of individual stars.

A related approach to calibrate masses and ages for giants using seismic data is presented in \cite{Ness2015b}. The Cannon confirms that the mass/age information is present in the APOGEE spectra, and that the five regions that carry most of the mass/age information correspond to four molecular CN lines and one molecular CO line. This is encouraging, and future papers will compare both methods of age determination: from the spectra with the Cannon, and from the element abundances with our techniques. We will also explore in more detail the implications of our work for the formation and evolution of the Milky Way by comparing the age structure of the Galaxy with numerical simulations.

\section*{Acknowledgments}
We thank the referee for suggestions that improved the presentation of the paper and the clarity of our arguments.
We thank Andrea Miglio and Nadege Lagarde for interesting discussions, as well as Friedrich Anders, Jo Bovy, Ricardo Carrera, and Matthew Shetrone for useful comments on this paper.
We thank Michele Cappellari for making his Voronoi binning code available in python. Analysis and plots presented in this paper used IPython and packages from NumPy, SciPy, Matplotlib, and Scikit-learn \citep{Hunter2007, Oliphant2007, Perez2007,Pedregosa2011}.
MM acknowledges support from the Alexander von Humboldt Foundation. The research has received funding from the European Research Council under the European Union's Seventh Framework Programme (FP 7) ERC Grant Agreement n. [321035]. SM has been supported by the J{\'a}nos Bolyai Research Scholarship of the Hungarian Academy of Sciences. 
D.A.G.H. and O.Z. acknowledge support provided by the Spanish Ministry of
Economy and Competitiveness under grant AYA-2014-58082-P. MP would like to acknowledge support from NASA grant NNX15AF13G. A.M.S. is partially supported by grants ESP2013-41268-R (MINECO) and 2014SGR-1458 (Generalitat of Catalunya).  
Funding for the Stellar Astrophysics Centre is provided by The Danish National Research Foundation (Grant agreement no.: DNRF106). The research is supported by the ASTERISK project (ASTERoseismic Investigations with SONG and Kepler) funded by the European Research Council (Grant agreement no.: 267864). V.S.A. acknowledges support from VILLUM FONDEN (research grant 10118).

Funding for SDSS-III has been provided by the Alfred P. Sloan Foundation, the Participating Institutions, the National Science Foundation, and the U.S. Department of Energy Office of Science. The SDSS-III Web site is http://www.sdss3.org/. SDSS-III is managed by the Astrophysical Research Consortium for the Participating Institutions of the SDSS-III Collaboration including the University of Arizona, the Brazilian Participation Group, Brookhaven National Laboratory, Carnegie Mellon University, University of Florida, the French Participation Group, the German Participation Group, Harvard University, the Instituto de Astrofisica de Canarias, the Michigan State/Notre Dame/JINA Participation Group, Johns Hopkins University, Lawrence Berkeley National Laboratory, Max Planck Institute for Astrophysics, Max Planck Institute for Extraterrestrial Physics, New Mexico State University, New York University, Ohio State University, Pennsylvania State University, University of Portsmouth, Princeton University, the Spanish Participation Group, University of Tokyo, University of Utah, Vanderbilt University, University of Virginia, University of Washington, and Yale University.

{}
\appendix 

\section{Fitting formulae and fit coefficients}\label{Appendix_fit}
In the following three tables we provide the best fit coefficients for the three different fits performed in the paper: mass as a quadratic function of first [M/H], [C/M], [N/M] and [(C+N)/M] (Table \ref{tab:coef_mass1}), and then [M/H], [C/M], [N/M],  [(C+N)/M], \teff and log $g$ (Table \ref{tab:coef_mass2}), and finally log(age) as a quadratic function of [M/H], [C/M], [N/M],  [(C+N)/M], \teff and log $g$ (Table \ref{tab:coef_age}).

As an example, to use Table \ref{tab:coef_mass1} to compute mass, one has to do the following:
$$
\mathrm{mass}=1.08-0.18\times\mathrm{[M/H]}-1.05 \times \mathrm{[M/H]}^2 \dots
$$
\begin{table*}
\begin{center}
\caption{Best fit coefficients for mass as a quadratic function of [M/H], [C/M], [N/M] and [(C+N)/M]}\label{tab:coef_mass1}
\begin{tabular}{cccccc}
 & 1 & [M/H] & [C/M] & [N/M]& [(C+N)/M] \\
\hline
 1   & 1.08   & -0.18   & 4.30  & 1.43   & -7.55 \\
$\mathrm{[M/H]}$  &   & -1.05   & -1.12  & -0.67   & -1.30 \\
$\mathrm{[C/M]}$ &  &   & -49.92   & -41.04   & 139.92 \\
$\mathrm{[N/M]}$& &  &    & -0.63   & 47.33 \\
$\mathrm{[(C+N)/M]}$ & & &   &  & -86.62 \\

\hline
\end{tabular}
\end{center}
\end{table*}

\begin{table*}
\begin{center}
\caption{Best fit coefficients for mass as a quadratic function of [M/H], [C/M], [N/M],  [(C+N)/M], \teff and log $g$}\label{tab:coef_mass2}
\begin{tabular}{cccccccc}
 & 1 & [M/H] & [C/M] & [N/M]& [(C+N)/M] & \teff/4000 & log $g$\\
\hline
1  & 95.87   & -10.40   & 41.36   & 15.05   & -67.61   & -144.18   & -9.42  \\
$\mathrm{[M/H]}$  &  & -0.73   & -5.32   & -0.93   & 7.05   & 5.12   & 1.52  \\
$\mathrm{[C/M]}$ &  &  & -46.78   & -30.52   & 133.58   & -73.77   & 16.04  \\
$\mathrm{[N/M]}$& &  &    & -1.61   & 38.94   & -15.29   & 1.35  \\
$\mathrm{[(C+N)/M]}$ & & &   &  & -88.99   & 101.75   & -18.65  \\
\teff/4000 & &  &  & &  & 27.77   & 28.80  \\
log $g$ & & &   & &  &  & -4.10  \\
\hline
\end{tabular}
\end{center}
\end{table*}

\begin{table*}
\begin{center}
\caption{Best fit coefficients for log(age) as a quadratic function of [M/H], [C/M], [N/M],  [(C+N)/M], \teff and log $g$}\label{tab:coef_age}
\begin{tabular}{cccccccc}
 & 1 & [M/H] & [C/M] & [N/M]& [(C+N)/M] & \teff/4000 & log $g$\\
\hline
1  & -54.35   & 6.53  & -19.02   & -12.18   & 37.22   & 59.58   & 16.14  \\
$\mathrm{[M/H]}$&   & 0.74   & 4.04  & 0.76   & -4.94   & -1.46   & -1.56  \\
$\mathrm{[C/M]}$&  &    & 26.90   & 13.33   & -77.84   & 48.29   & -13.12  \\
$\mathrm{[N/M]}$& & &    & -1.04   & -17.60   & 13.99   & -1.77  \\
$\mathrm{[(C+N)/M]}$&  & & &    & 51.24  & -65.67   & 14.24  \\
\teff/4000 & &  &  & &  & 15.54   & -34.68  \\
log $g$ & & &   & &  &    & 4.17  \\
\hline
\end{tabular}
\end{center}
\end{table*}

\end{document}